\documentclass{article}

\usepackage[final]{neurips_2022}

\usepackage[utf8]{inputenc} 
\usepackage[T1]{fontenc}    
\usepackage{hyperref}       
\usepackage{url}            
\usepackage{booktabs}       
\usepackage{amsfonts}       
\usepackage{nicefrac}       
\usepackage{microtype}      
\usepackage{xcolor}         
\usepackage{graphicx}

\usepackage{todonotes}

\title{BigScience: A Case Study in the Social Construction \\ of a Multilingual Large Language Model}

\author{%
  Christopher Akiki \\
  Leipzig University \\
  \And
  Giada Pistilli \\
  Hugging Face \\
  \And
  Margot Mieskes \\
  Hochschule Darmstadt
  \AND
  Matthias Gallé \\
  Cohere \\
  \And
  Thomas Wolf \\
  Hugging Face \\
    \And
  Suzana Ilić \\
  MLT \\
  \And
  Yacine Jernite \\
  Hugging Face
}
\begin{document}
\maketitle
\begin{abstract}
The BigScience Workshop was a value-driven initiative that spanned one and half years of interdisciplinary research and culminated in the creation of ROOTS, a 1.6TB multilingual dataset that was used to train BLOOM, one of the largest multilingual language models to date.
In addition to the technical outcomes and artifacts, the workshop fostered multidisciplinary collaborations around large models, datasets, and their analysis. 
This in turn led to a wide range of research publications spanning topics from ethics to law, data governance, modeling choices and distributed training. 
This paper focuses on the collaborative research aspects of BigScience and takes a step back to look at the challenges of large-scale participatory research, with respect to participant diversity and the tasks required to successfully carry out such a project. 
Our main goal is to share the lessons we learned from this experience, what we could have done better and what we did well. We show how the impact of such a social approach to scientific research goes well beyond the technical artifacts that were the basis of its inception.
\end{abstract}

\section{BigScience Workshop---Context and Inception}
\label{introduction}
Research practices are inevitably tied to the socio-technical contexts in which they are embedded. Such a contextual and fluid view is, according to \citet{kuhn:1962}, part and parcel of the scientific enterprise, whose necessary evolution is modulated by revolutions leading to new paradigms. A particularly useful paradigmatic view of the scientific method as it relates to---and is transformed by---computing technologies can be found in Jim Gray's last talk he gave before disappearing at sea and the posthumous anthology~\citep{hey2009-fourth-paradigm} it inspired. \citeauthor{hey2009-fourth-paradigm} saw in the commodification of data a transformation of how research is conducted. \citet{Symons2014-software-intensive} characterized this mode of data-driven research as primarily software-intensive, a characterization that is especially true for modern deep learning \citep{bekman-2022-accelerate,bekman-2022-bloom}, making meaningful research contingent upon the formation of more specialized teams; a need that would---among other things---also come to characterize \textbf{``Big Science''}: a specific form science that emerged in the 1940s~\citep{sep-longino-scientific-knowledge-social}.

\begin{figure}[ht]
\centering
\includegraphics[width=\textwidth]{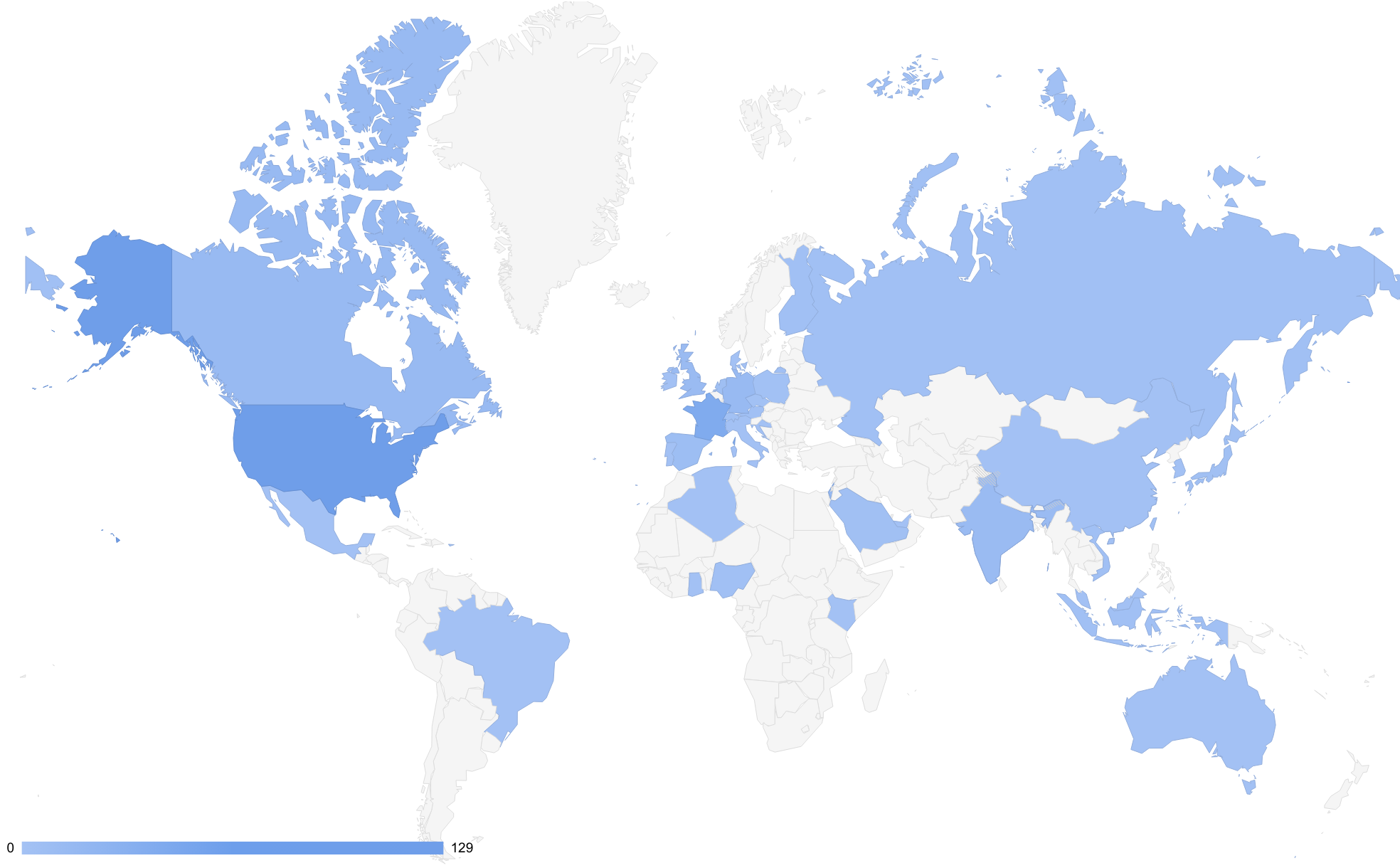}
\caption{Geographic location of residence for 308 BigScience participants with a at least one traced contribution. This corresponds to 38 countries. (See Section~\ref{effective-diversity} for more information.)}
\label{fig:locations}
\end{figure}

This Big~Science phenomenon grew out of the necessity to cope with the increasing complexity of twentieth century research questions and agendas. Thousands of researchers of diverse backgrounds and expertise, organized in specialized sub-groups, have on various occasions collaborated together over extended periods of time to be able to achieve what no individual effort could possibly hope to manage: land on the moon~\citep{apollo}, accurately estimate the mass of the Higgs Boson~\citep{higgs-boson}, sequence the human genome~\citep{human-genome}, and detect gravitational waves~\citep{ligo}. It was indeed this sort of large-scale multidisciplinary collaboration that inspired the creation of the BigScience Workshop.

The BigScience Workshop project originated from discussions in late 2020 and early 2021 between Thomas Wolf (Hugging Face), Stéphane Requena (GENCI) and Pierre-François Lavallée (IDRIS); GENCI and IDRIS being respectively the designer--builder and operator of the French supercomputer ``Jean Zay'', a national computing center for the CNRS ("Centre national de la recherche scientifique", the French National Research Organization). These early discussions went over the possibilities that a large cluster like Jean Zay with close to 2700 GPUs could offer to the field of Artificial Intelligence. 
Quickly this converged toward the goal of training a very large language models, of the order of 100 billions of parameters.
With respect to existing such models, the identified issues was that most of these models are currently trained privately with no oversight from the research community at large, but more crucially the people at the receiving end of these technologies who stand to be hurt the most by them. 

A popular belief---fueled by the commodification of data---is that data is a mere value-less true representation of the world and therefore a ``harbinger of transparency, democracy and social equality~\citep{sep-leonelli-science-big-data}.
In reality however, the \textbf{digital divide}~\citep{sullins-divide} often  extends naturally into a \textbf{data divide} which inherently limits the representativeness of any data, owing to the ever-widening gap between those who can access ICT~(Information and communications technology) infrastructure and those who cannot. This absence of data relating to certain socioeconomic, socio-cultural, and geographic groups inherently limit the comprehensiveness of any data resource~\citep{sep-leonelli-science-big-data} and renders any artifact that builds on such data---such as language models---into a tool that reinforces and potentially amplifies the inequalities encoded in large datasets~\citep{stochastic}.

Unfortunately, this commodification of data could in practice lead to an unreflected leveraging of the Web as a convenient source of large quantities of training material~\citep{birhane-laion}, especially by companies whose identity is ``strongly linked with data''~\citep{beaulieu-leonelli-2021-data} who have an incentive to default to what \citet{krohs-convenience-experimentation} calls \textbf{convenience experimentation}---that is experimental designs, practices, methods, and data that are adopted not because of their suitability to the problem at hand, but because they are ``easily and widely available and usable, and thus convenient means''~\citep{sep-leonelli-science-big-data} for private research labs to achieve their goals. 

Being cognizant of these challenges, the BigScience Workshop adopted a value-driven~\citep{elliott-tapestry-values} approach, grounded in an ethical charter~(See Section~\ref{sec:ethics-governance}), that modulated all processes involved in the training of the BLOOM~model\footnote{\url{https://hf.co/bigscience/bloom}}, the creation of the ROOTS~corpus~\citep{roots}, and all other workshop outputs~(See Section~\ref{sec:outputs-lessons-future}). Targeted diversity (See Sections~\ref{sec:ethics-governance}~and~\ref{effective-diversity})---both socio-cultural and disciplinary---was a key ingredient in the success of the workshop. The benefits of such an inclusive and diverse participatory approach to research, what \citet{birhane-participatory} call the ``participatory turn'' of AI research, goes well beyond the Big Science metaphor and is indeed well aligned with trends observed by \citet{wang_barabási_2021}, who attempt to attempt to quantify the effects of the institutionalization of 20th century science~\citep{sep-longino-scientific-knowledge-social}, and use publication data to observe the 1)~growing importance of teams across disciplines, 2)~the internationalization of research collaborations, 3)~the importance of diversity---ethnic, geographic, and institutional---and its positive effect on scientific impact, and 4)~the importance of the research dynamics of big teams in knowledge-production~\citep{wang_barabási_2021}. This shows the importance of community-driven collaborative ML and AI collectives~\citep{the_turing_way_community_2022_6909298} and explains their recent proliferation and positive impact on the field. Non-profit social-participation collectives such as EleutherAI, the ML Collective, Cohere for AI, MLT, Masakhane, MD4SG, and BigScience form an important counterweight to a field that often relegates issues of ethics, harm, and governance to secondary positions of post-facto crisis management and damage control. This ``train first, ask questions later'' approach to AI was exactly what the BigScience Workshop attempted to avoid, and what this current paper attempts to elucidate.

\section{Value-Driven Science: Organization, Governance, and Participation}
\label{sec:ethics-governance}

The BigScience project was initiated in January 2021, a few months after \citet{stochastic} brought attention to the risks inherent in the approach of prioritizing increasing model size as the main path forward to ``improving'' Machine Learning systems. It also followed recent calls to further examine the values encoded both in the datasets that support ML research and in the research practices themselves \citep{dataset-politics,values-ml}. In this context, and in order to start addressing some of the limitations outlined in these works, the BigScience project started as a request for a large compute grant on the French public supercomputer Jean Zay~\footnote{\href{http://www.idris.fr/eng/jean-zay/jean-zay-presentation-eng.html}{Jean Zay supercomputer}} that would allow a greater range of participants (especially outside of the best-resourced US-based industrial lab) to work on defining, developing, and interrogating a Large Language Model of a similar size to ones recently developed \citep{gpt3}. In particular, the grant request~\footnote{Available \href{https://drive.google.com/file/d/1l-hKP2lFIvvcqpMryuD5GVYOOBqlubA_/view}{here}.} emphasized \textbf{openness}, \textbf{inclusion}, and \textbf{responsibility} as driving values for the project.

In order to meet these objectives, we first endeavored to \textbf{map research topics} that were relevant to fostering these values in the development of LLMs, and to set up a \textbf{project organization and governance structure} focused on enabling an open distributed collaboration driven by shared values while fostering diverse participation.

\subsection{Mapping Research Topics}
\label{sec:mapping-research}

The BigScience workshop was devised as an open research collaboration organized around the production of a specific artifact: a multilingual Large Language Model to be made available to the ML research community to support further investigation. The creation of such an artifact raised a number of interdependent but distinct research questions, especially for a project that aimed to meaningfully engage with its social context and acknowledge its social dimensions~\citep{artifact-politics}.

\begin{figure}[ht]
\centering
\includegraphics[width=\textwidth]{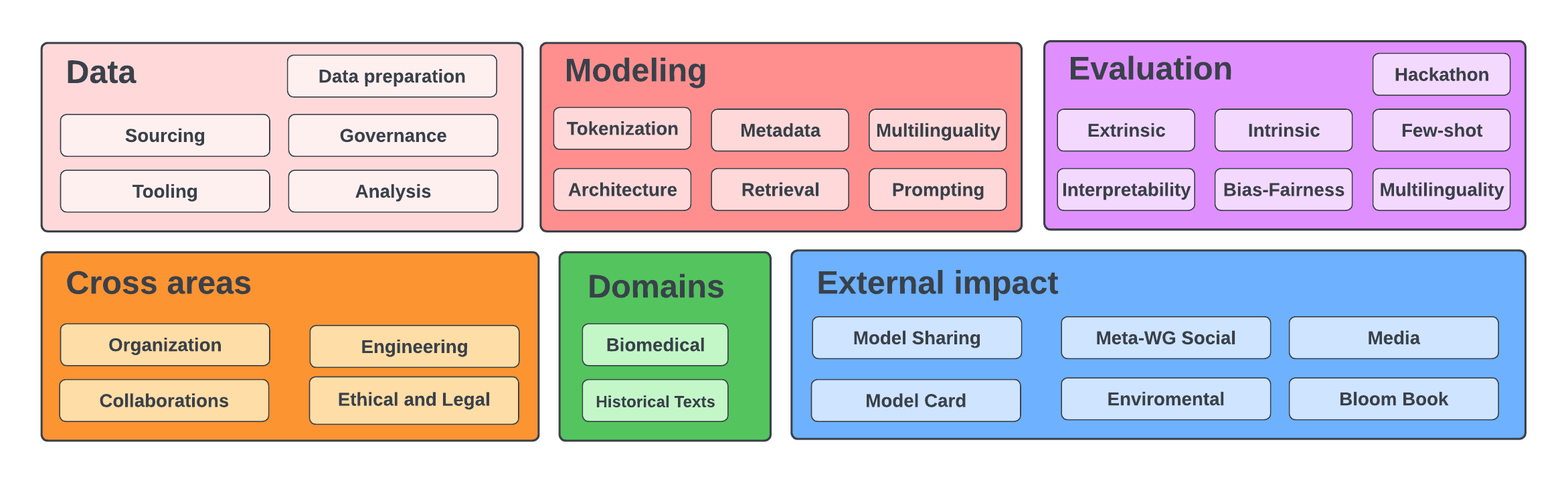}
\caption{The BigScience working groups}
\label{fig:workinggroups}
\end{figure}

This network of related research questions was reflected in the project's organization into Working Groups. Each Working Group comprised several participants with various levels of involvement including a few chairs whose role was to self-organize around a specific aspect of the overall project. Importantly, participants were encouraged to join more than one working group in order to share experiences and information. During the preparatory phase of the project launching up to the May 2021 launch event, we defined a starting set of working groups corresponding to the initial expertise and interests of the participants.\footnote{\href{https://docs.google.com/presentation/d/1ITOEHnVcfXuRfooi7WOh5j3xl742o-cDK0AjoFXw5_g/edit\#slide=id.gd36cc9732b_0_0}{List of Working Group categories} at the launch event.} We also invited participants to start new working groups as the need arose and as the diversity of the expertise and experience in the workshop increased.
Indeed, the 10~initial proposals grew into the set of 30~working groups presented in Figure~\ref{fig:workinggroups}.

The choice of which research questions to prioritize is significant for a project of this size. The behavior of the final trained multilingual model that was the focus of the effort would to the best of our knowledge depend on a range of \textbf{Modeling} choices, including for example tokenization~\citep{korean-tokenization} or architecture~\citep{bigscience-architecture}; all of which were explored in specific modeling working groups. Training a very large model on a cluster like Jean Zay also presents unique and novel challenges, which were addressed by the members of the \textbf{Engineering} working group.~\footnote{\href{https://github.com/huggingface/blog/blob/main/bloom-megatron-deepspeed.md}{Overview} of the Engineering WG.} These working groups together aimed to ensure that the best possible use was made of the consequent compute resources made available by the grant supporting the project.

The project was also motivated by a drive to better \textit{understand} trained Large Language Models. Thus, being able to properly evaluate various aspects of the model's behavior was instrumental both to measuring the impact of choices made during the project and to furthering the community's understanding of this category of systems' general properties. BigScience's various \textbf{Evaluation} working groups worked on adapting recent notable efforts to develop evaluation suites for LLMs~\citep{eval-harness,bigbench} and extending their scopes to more languages, exploration and visualization tools, and evaluation methods.

A trained LLM is also a reflection of its training \textbf{Data}. Recent work has drawn attention to various issues caused by the lack of value put on data work in our research community~\citep{data-cascades}, and to how prioritizing efficiency and technical performance comes at the expense of social considerations for datasets~\citep{dataset-politics,laion-analysis}; including over-relying on automatic curation that fails to examine the additional biases it introduces~\citep{c4-analysis}. In contrast, we made data elicitation and curation a significant part of our effort, with groups dedicated to questions of sourcing, governance, preparation, analysis, and other necessary tooling. This made it easier to intentionally select what language would be included in the final corpus and to foster diversity and awareness of the data subjects.

Considerations of \textbf{Social Impact and Context} were spread across the whole projects, including but not limited to the data governance working group mentioned above, an evaluation working group focused on fairness evaluation, work on the carbon footprint of the project, etc. Among those, the \textbf{Ethical and Legal Scholarship} played a special role by laying the foundation for broader, collaborative work among the different working groups in a horizontal and participatory effort.
Through their complementarity, the philosophical and legal disciplines guided the framework for the governance of BigScience’s artifacts, thus laying the foundation for broader discussion.
The most visible outcomes of this work were a project-wide ethical charter~\footnote{\href{https://bigscience.huggingface.co/blog/bigscience-ethical-charter}{BigScience Ethical Charter}.}, a model Responsible AI License to account for downstream uses of the model~\footnote{BigScience \href{https://bigscience.huggingface.co/blog/bigscience-openrail-m}{OpenRAIL-M License}.}, and a week-long legal hackathon where 30 legal scholars investigated the international legal context for the technology~\footnote{BigScience \href{https://bigscience.huggingface.co/blog/legal-playbook-for-natural-language-processing-researchers}{Legal Playbook}.}.

Finally, the success of the overall project was highly dependent on the work of the \textbf{Organization and Communication} working groups whose missions included fostering cross-group communication, organizing regular events that served as milestones for the full community of participants --- including the closing workshop at ACL 2022 ---, and managing the logistics of the project to allow new participants to easily join and existing participants to keep abreast of the many ongoing efforts.

\subsection{Distributed Project Organization, Governance, and Diversity}

\paragraph{Workshop Organization and Communication Mechanisms}
The BigScience Workshop used multiple communication channels for communication and organization. Most of the discussions happened on the \href{https://huggingface.co/}{Hugging Face} company Slack, where participants were invited to join as multi-channel guests with access to the channels corresponding to the working groups they had joined. For the sake of visibility, all working documents were hosted on a \href{https://drive.google.com/drive/u/1/folders/1db2hYZuRs2VjoIrVaVtZJ5FLE2iS7z3p}{Google Drive folder} which by default had universal read access, and write access for members of the specific working groups. The comment threads on the documents in this drive were also an important channel of communication. Regular synchronizations were also organized in the form of project-wide live events (6 including the \href{https://bigscience.notion.site/Kick-off-event-April-28th-2021-5f3d29cb898846e8916ff700662c3319}{kick-off} and \href{https://bigscience.huggingface.co/acl-2022}{closing workshop}) and more frequent bi-weekly calls between all the working group chairs, as well as a regular newsletter sent out to all participants. Finally, many of the project participants came from an open-source software (OSS) culture, and many of the project's contributions came in the form of open-source software, so a significant portion of the conversations and many of the technical decisions were taken through GitHub interactions~\footnote{Github \href{https://github.com/bigscience-workshop/}{BigScience organization}.} (discussions and pull requests).

The communication approach was designed with an aim to foster inclusion by putting asynchronous written communication first, and enabling a consensus-based decision mechanism where all concerns from participants directly affected or with expertise relevant to a decision were addressed before moving forward. In particular, the chairs were asked to coordinate between working groups to ensure that people across the organization were aware when decisions that were relevant to them were being discussed. In practice, however, we still found that live meetings were instrumental in communicating more nuanced information, but could be particularly difficult with participants on all continents.

An additional challenge came from the project's somewhat restricted time frame. Many of the different research topics outlined in Section~\ref{sec:mapping-research} depend on each other. For example, focusing only on the data aspect of the work, having a good grasp of data governance processes should precede working to identify data sources, which needs to be done before the data is prepared and then, analyzed; an analysis which should then again inform new governance practices. In particular, in most of these cases, the sharp increase in scale in the last two years makes it difficult to rely on existing work. However, as we were strictly constrained by the availability of the computing resources that would be used to train the model and put a time limit on when the training corpus should be available, we had to do our best to do as much of this work in parallel, with more or less success depending on the aspects.

\paragraph{Aligning Goals through an Ethical Charter}

One way to empower our diversity has been to use an appropriate normative ethics framework to let coexist and enhance our scientific, cultural, and professional diversity. Through the adoption of a value pluralist approach~\citep{handbook-value}, according to which the order of moral values may vary but cannot be considered less important, we framed our method. The best way to make this approach work is to inscribe it in a principle belonging to the Confucian moral theory tradition: the principle of harmony~\citep{harmony}. 

Once the scope of action and normative approach had been defined, we started drafting the \href{https://bigscience.huggingface.co/blog/bigscience-ethical-charter}{ethical charter}, which aims to engage us individually and collectively. So the need to have an ethical charter stems from an awareness of the possible negative repercussions associated with the development of LLMs (as stated in the charter's preamble) but at the same time, a willingness to commit on a moral level to defined and shared values. These same values were later reused and developed vertically by the different WGs working on specific issues with particular ethical challenges. Added to the approach described above is the distinction between intrinsic and extrinsic values ~\citep{handbook-intrinsic} that we have adopted. This value theory allowed us to have the agility to represent pivotal, intrinsic values as unshakable and long-lasting over time. We refer here, for example, to the value of inclusivity: described as a sense of belonging and feeling welcome, it becomes an enduring value within the BigScience project. On the other hand, extrinsic and thus instrumental values achieve the goals set by intrinsic ones and can be replaced over time. In our example, the extrinsic value of interdisciplinarity becomes essential in order to achieve the intrinsic value of inclusivity: the two become essential to each other.

Writing the ethical charter as a collaborative and consensus-based endeavor presented particular challenges.
First, moral emotions \citep{moral-emotions} came into play when we had to discuss definitions of BigScience values, that is, those social emotions that animate conversations about what we care about. This made alternating between bi-weekly live meetings to channel these discussions with periods of asynchronous written exchanges (between Slack and document comments) particularly important. 
Second, getting participation from the greatest number of project collaborators required significant effort. Engagement increased after the first draft, which allowed us to have a more solid basis for discussion.
The limitations of non-physical collaboration with participants in the same project were evident there, but the challenge allowed us to get creative. For instance, adopting the latest version of the ethical charter was done through a questionnaire; while it left less room for nuanced discussion of the individual points, it made it possible to reach those collaborators who did not have time to engage in ethical discussions.


\paragraph{Building Diversity}
The BigScience workshop aimed to increase the range of expertise and experiences who take part in shaping new technology, and to promote the agency of under-represented voices in doing so. It also strove to be cognizant of ways in which attempts to foster diversity without interrogating for whose benefit can run contrary to this goal. While improving the representation of non-European languages in NLP technology can be a worthy goal~\citep{nlp-ling-diversity}, attempts to develop resources under the full direction and ownership of a handful of institutions outside of their context become extractive ``helicopter research''~\citep{helicopter-research}. Recent scholarship has also explored how traditional discourses of inclusion can reinforce harmful frames and paradigms~\citep{terms-of-inclusion} and how the disproportionate role of technology companies in social impact research can hobble efforts in that space~\citep{facct-corporate-capture}.
In addition to fostering an inclusive environment via its consensus-based organization, \href{https://bigscience.huggingface.co/blog/bigscience-ethical-charter}{ethical charter}, and \href{https://docs.google.com/document/d/14Q_wtc7QGdQ0wIBK3zQbCblQn84tfaD3/}{code of conduct}, the BigScience workshop strove to address the pitfalls outlined above by focusing specifically on increasing agency in our outreach efforts.

The first priority to that end was to reach out to potential participants outside of our immediate networks early in the project \textbf{while the goals and approach were still being defined}. We started by identifying partner organizations (primarily grassroots organization, advocacy groups focused on internet and equity, national libraries, and universities with at least one faculty member working on NLP) based on criteria of geographical diversity and expertise in relevant fields, including sociolinguistics, technology regulation, and technology governance. We found that most people we reached out to on that basis with a high-level explanation of the overall workshop goals and \textit{where we thought their specific expertise would fit in the project} were willing to schedule a video call for further information, and to direct us to some of their colleagues who might be a better fit when they themselves could not join the project. Secondly, we put an emphasis on \textbf{diversity in leadership positions} as much as on the diversity of overall participants. The organization group in particular worked to that end by reaching out to individual participants and collecting feedback on what would make it easier for them to serve as chairs.

Last but not least, we endeavored to make the BigScience workshop \textbf{inclusive to research that did not directly contribute to the final artifacts}. The goal was again to give participants the flexibility to define how they could best benefit from their own work within BigScience, and foster a mutually beneficial partnership rather than a one-way transfer of skills. This led for example to working groups that branched off as their own projects, such as the efforts focused on \href{https://github.com/bigscience-workshop/biomedical}{biomedical data} and \href{https://github.com/bigscience-workshop/lam}{historical text}. It also informed how we ran e.g. data sourcing hackathons~\citep{bigscience-catalog} where participants were asked to index language resources that were of broad interest to their work not restricted by their fitness to our specific use case to make the resulting catalog useful beyond BigScience.

\section{BigScience Participants Post Hoc Diversity and Feedback Survey}
\label{effective-diversity}


Of the over 1200 people registered to BigScience and were given access to its communication channels, we found that 365 individuals had directly contributed to the project's released artifacts in a way that we could trace. It is important to note that while the largest group originated from the US, almost all continents were represented in the project, ranging from Asia, Africa, North and South America and Europe as can be seen on the map in Figure~\ref{fig:locations}---a total of 38 countries: China, Japan, Taiwan, Hong Kong, Vietnam, Indonesia, Singapore, Malaysia, India, Saudi Arabia, United Arabic Emirates, Israel, Kenya, Nigeria, Ghana, Portugal, Spain, France, Germany, Czech Republic, Poland, Denmark, Netherlands, Finland, Russia, Canada, Mexico, Puerto Rico, and Brazil\footnote{These are countries of residence, not origin.}.

At the conclusion of the BigScience project, we also carried out a survey among the participants. While only 24 answered, the answers give an interesting insight into various other aspects of the collaboration within the project.
The following information is drawn from this survey, which contained various questions, ranging from demographic questions to open questions, where participants could express their opinion freely and openly.
The results of this survey also support the cultural diversity among the participants.
But it also showed that their background is just as diverse. While the majority comes from a computer science background, a lot of participants had an additional background in for example linguistics, statistics, socio-cultural antrhopology or law.
Few participants had a non-CS background, such as philosophy or law. 
This resulted in also quite homogeneous working groups, where most people stated that they were collaborating with other computer scientists. 
But some stated that they collaborated with people with law, philosophy, ethics, sociology or GLAM background -- probably also depending on the actual working groups.

Nevertheless, in general the communication within the groups was rated very positively, while the communication across the various working groups was rated a lot lower -- so this would be something to improve in another, similar project. 
But, across the whole project, the collaboration was rated quite highly. 
Also the languages represented were quite diverse -- as could be expected from a project that aims to build a multilingual language model. 
English was the dominating language, followed by German, French, Spanish and Arabic, but lower resource languages such as Norwegian or Niger Congo languages were also worked on. 
The majority of participants joined the project on a voluntary basis, without being explicitly paid to do so. Most did so, because they wanted to learn something or because they believed in the overall goals of the project. 
The project as a whole was rated very high and when asked about the achieved goals, most answers indicated that almost all goals were achieved, even if not perfectly and some issues were still open at the time of writing.
Overall, participants liked the openness of the project and the community as a whole, which is described as inclusive and multicultural. 
Things participants expressed a dislike on, was various factors, such as the communication across groups, or finding your footing if one joined later in the project, as there were so many channels, so many groups and things grew organically throughout the project. 
Also the dominance of English was criticized, but it might be difficult to change that. 
When it comes to doing things differently in the future, most answers asked for a bit more steering, having the possibility to join earlier and more funding.
At the end, nobody expressed that they would not join a follow-up project, on the contrary, almost 70\% of the participants indicated, that they would participate in a follow-up project.

\section{Lessons Learned, Workshop Outputs, and the Future of BigScience}
\label{sec:outputs-lessons-future}
If an end-date has to be put to this initiative, it could be the last (hybrid) workshop~\citep{bigscience-2022-bigscience}, on May 27th 2022\footnote{\url{https://bigscience.huggingface.co/acl-2022}}. 
While this concluded the more organized efforts, several working groups continued either wrapping up or even brainstorming new ideas.
In particular, the model\footnote{\url{https://hf.co/bigscience/bloom}} (dubbed BLOOM) was released in early July.

When reflecting back on this endeavour, we believe that it showed the possibility of setting up a (very) large collaborative structure in the area of machine learning, something which to our knowledge had not been done at this scale before.
We argue that part of is success can be attributed to a very conscious effort to encompass the global community. 
This is true both at the geographical sense, as well as skill-wise: the BigScience included not only researchers  with technical background in training large language models, but also ethicists, social scientists, legal scholars, and practitioners. 
Beyond the final model, BigScience created a large list of papers 
and spurred new collaborations, often between people who would not have met otherwise.
More generally, it showed that open collaborations can work not only for small paper-like projects, but also for more ambitious projects that require various intermediate steps.
Beyond ROOTS and BLOOM, this initiative spawned at least 16 papers\footnote{See complete list \href{https://github.com/bigscience-workshop/bibliography/blob/master/bigscience.bib}{here}} and several other assets not necessarily (yet) described in a research paper.
Those include a consortium focusing on multi-modal (speech+text) models funded by the European Commission; as well as the follow-up project BigCode\footnote{\url{https://www.bigcode-project.org/}} and BigLAM\footnote{\url{https://github.com/bigscience-workshop/lam}} which were launched very recently.

In order to best meet its goals, the BigScience project involved a number of trade-offs which---in hindsight---could have been better negotiated to make for a smoother experience.

\paragraph{Legal entity or ad-hoc collaboration.}
One of the questions that came up at different points during the project was whether it was better run as an informal collaboration between individual volunteers (with support from the host organization Hugging Face), or whether it would be its own legal entity, possibly with the capacity to raise proper funds and hire staff. We ended up remaining in the former situation for the length of the project, not least because the latter would have taken too long to set up given the overall timeline. Having an informal collaboration made it easier for participants to join without too much oversight from their main employers, especially participants whose main position was in industry. Requiring them to get formal approval from their management chain to join e.g. an established consortium would have been significantly more cumbersome and might have proved detrimental to the general enthusiasm for the project.

At the same time, this lack of legal entity made it more complicated to join for those companies whose legal department had a strong say in internal decisions and employee activities. There was also no way for contributors to get remunerated for their work, or funds for expenses outside of compute (e.g., licensing fees).
Individuals participated because they believed in the vision of the project, and/or because of some expected follow-up gain (visibility, employment opportunities, co-authoring some assets, training possibilities,  networking, etc). 
This made every effort dependent on this intrinsic motivation of each individual, as well as timing commitment outside pressing deadlines of other responsibilities they might have. 
More generally, the project was from the beginning very bottom-up and consensus based.
The associated difficulties with that and the need of taking decisions and fulfilling some milestones at concrete deadlines was often solved by the initial institutions (Hugging Face) dedicating some resources to solve that problem. 
It is far from certain that the project would have accomplished what it did without those dedicated resources. 

\paragraph{Breadth, time, and participation.} Defining the scope of the project was another challenge. The minimal goal of "training a multilingual large language model" could have been achieved with significantly fewer participants; some of the modeling working groups, the engineering working group, and the work needed to filter an existing data source such as the OSCAR corpus~\citep{oscar}. This would not, however, have met the project's goal of responsibility and inclusivity that were the motivation for the approach. Addressing various social and technical aspects of LLMs together also provided a rare opportunity for scholars from different discipline to interact directly and work on problems that require diverse expertise. On the other hand, the more interdependent aspects of LLMs we aimed to address together, the harder it became to plan project steps, since some of the work did have to happen in sequence. This particular challenge came in great part from the novelty of the approach, and the original uncertainty about how many people, and with what expertise, would be interested in joining; we hope future endeavors of this kind will be able to better scope the research areas and dependency graphs between their outputs further ahead.

\paragraph{Flexible goals and planning ahead.} Relatedly, while
flexibility in both the project structure and the framing of its output was necessary to foster true inclusion and take action based on feedback from our diverse participants, it did make overall project planning that much more difficult. Doing so would have been even harder without the support of the two Hugging Face employees who worked as full-time and part-time Technical Program Managers respectively, and we strongly recommend future projects dedicate significant resources to these roles early on.

The BigScience Workshop presented a novel way of collaborating on large-scale ML models that aimed to prioritize foresight and breadth of expertise. In addition to the direct outcomes of the project, we hope it will provide a blueprint, or at least an inspiration for future endeavors that want to do better than the ``train first, ask questions later'' approach we have seen in recent years; and foster a more inclusive and thoughtful development of ML technology.




\begin{ack}
The BigScience Workshop was granted access to the HPC resources of the Institut du développement et des ressources en informatique scientifique (IDRIS) du Centre national de la recherche scientifique (CNRS) under the allocation 2021-A0101012475 made by Grand équipement national de calcul intensif (GENCI). Model training ran on the Jean-Zay cluster of IDRIS, and we thank the IDRIS team for their responsive support throughout the project, in particular Rémi Lacroix.
\end{ack}

\bibliographystyle{chicago}
\bibliography{bibliography}

\begin{thebibliography}{}

\bibitem[\protect\citeauthoryear{Aad, Abbott, Abdallah, Das, Di~Giovanni,
  Field, and Fisher}{Aad et~al.}{2015}]{higgs-boson}
Aad, G., B.~Abbott, D.~Abdallah, J.~Curry, S.~Das, G.~P. Di~Giovanni, R.~D.
  Field, and M.~Fisher (2015, May).
\newblock Combined measurement of the higgs boson mass in $pp$ collisions at
  $\sqrt{s}=7$ and 8 tev with the atlas and cms experiments.
\newblock {\em Phys. Rev. Lett.\/}~{\em 114}, 191803.

\bibitem[\protect\citeauthoryear{{Abadji}, {Ortiz Suarez}, {Romary}, and
  {Sagot}}{{Abadji} et~al.}{2022}]{oscar}
{Abadji}, J., P.~{Ortiz Suarez}, L.~{Romary}, and B.~{Sagot} (2022, January).
\newblock {Towards a Cleaner Document-Oriented Multilingual Crawled Corpus}.
\newblock {\em arXiv e-prints\/}.

\bibitem[\protect\citeauthoryear{Abbott, Abbott, Abbott, Abernathy, Acernese,
  Ackley, Adams, Adams, Addesso, Adhikari, Adya, Affeldt, Agathos, Agatsuma,
  Aggarwal, Aguiar, Aiello, Ain, Ajith, Allen, Allocca, Altin, Anderson,
  Anderson, Arai, Arain, Araya, Arceneaux, Areeda, Arnaud, Arun, Ascenzi,
  Ashton, Ast, Aston, Astone, Aufmuth, Aulbert, Babak, Bacon, Bader, Baker,
  Baldaccini, Ballardin, Ballmer, Barayoga, Barclay, Barish, Barker, Barone,
  Barr, Barsotti, Barsuglia, Barta, Bartlett, Barton, Bartos, Bassiri, Basti,
  Batch, Baune, Bavigadda, Bazzan, Behnke, Bejger, Belczynski, Bell, Bell,
  Berger, Bergman, Bergmann, Berry, Bersanetti, Bertolini, Betzwieser, Bhagwat,
  Bhandare, Bilenko, Billingsley, Birch, Birney, Birnholtz, Biscans, Bisht,
  Bitossi, Biwer, Bizouard, Blackburn, Blair, Blair, Blair, Bloemen, Bock,
  Bodiya, Boer, Bogaert, Bogan, Bohe, Bojtos, Bond, Bondu, Bonnand, Boom, Bork,
  Boschi, Bose, Bouffanais, Bozzi, Bradaschia, Brady, Braginsky, Branchesi,
  Brau, Briant, Brillet, Brinkmann, Brisson, Brockill, Brooks, Brown, Brown,
  Brown, Buchanan, Buikema, Bulik, Bulten, Buonanno, Buskulic, Buy, Byer,
  Cabero, Cadonati, Cagnoli, Cahillane, Bustillo, Callister, Calloni, Camp,
  Cannon, Cao, Capano, Capocasa, Carbognani, Caride, Diaz, Casentini, Caudill,
  Cavagli\`a, Cavalier, Cavalieri, Cella, Cepeda, Baiardi, Cerretani, Cesarini,
  Chakraborty, Chalermsongsak, Chamberlin, Chan, Chao, Charlton,
  Chassande-Mottin, Chen, Chen, Cheng, Chincarini, Chiummo, Cho, Cho, Chow,
  Christensen, Chu, Chua, Chung, Ciani, Clara, Clark, Cleva, Coccia, Cohadon,
  Colla, Collette, Cominsky, Constancio, Conte, Conti, Cook, Corbitt, Cornish,
  Corsi, Cortese, Costa, Coughlin, Coughlin, Coulon, Countryman, Couvares,
  Cowan, Coward, Cowart, Coyne, Coyne, Craig, Creighton, Creighton, Cripe,
  Crowder, Cruise, Cumming, Cunningham, Cuoco, Canton, Danilishin, D'Antonio,
  Danzmann, Darman, Da~Silva~Costa, Dattilo, Dave, Daveloza, Davier, Davies,
  Daw, Day, De, DeBra, Debreczeni, Degallaix, De~Laurentis, Del\'eglise,
  Del~Pozzo, Denker, Dent, Dereli, Dergachev, DeRosa, De~Rosa, DeSalvo,
  Dhurandhar, D\'{\i}az, Di~Fiore, Di~Giovanni, Di~Lieto, Di~Pace, Di~Palma,
  Di~Virgilio, Dojcinoski, Dolique, Donovan, Dooley, Doravari, Douglas, Downes,
  Drago, Drever, Driggers, Du, Ducrot, Dwyer, Edo, Edwards, Effler, Eggenstein,
  Ehrens, Eichholz, Eikenberry, Engels, Essick, Etzel, Evans, Evans, Everett,
  Factourovich, Fafone, Fair, Fairhurst, Fan, Fang, Farinon, Farr, Farr,
  Favata, Fays, Fehrmann, Fejer, Feldbaum, Ferrante, Ferreira, Ferrini,
  Fidecaro, Finn, Fiori, Fiorucci, Fisher, Flaminio, Fletcher, Fong, Fournier,
  Franco, Frasca, Frasconi, Frede, Frei, Freise, Frey, Frey, Fricke, Fritschel,
  Frolov, Fulda, Fyffe, Gabbard, Gair, Gammaitoni, Gaonkar, Garufi, Gatto,
  Gaur, Gehrels, Gemme, Gendre, Genin, Gennai, George, Gergely, Germain, Ghosh,
  Ghosh, Ghosh, Giaime, Giardina, Giazotto, Gill, Glaefke, Gleason, Goetz,
  Goetz, Gondan, Gonz\'alez, Castro, Gopakumar, Gordon, Gorodetsky, Gossan,
  Gosselin, Gouaty, Graef, Graff, Granata, Grant, Gras, Gray, Greco, Green,
  Greenhalgh, Groot, Grote, Grunewald, Guidi, Guo, Gupta, Gupta, Gushwa,
  Gustafson, Gustafson, Hacker, Hall, Hall, Hammond, Haney, Hanke, Hanks,
  Hanna, Hannam, Hanson, Hardwick, Harms, Harry, Harry, Hart, Hartman, Haster,
  Haughian, Healy, Heefner, Heidmann, Heintze, Heinzel, Heitmann, Hello,
  Hemming, Hendry, Heng, Hennig, Heptonstall, Heurs, Hild, Hoak, Hodge, Hofman,
  Hollitt, Holt, Holz, Hopkins, Hosken, Hough, Houston, Howell, Hu, Huang,
  Huerta, Huet, Hughey, Husa, Huttner, Huynh-Dinh, Idrisy, Indik, Ingram, Inta,
  Isa, Isac, Isi, Islas, Isogai, Iyer, Izumi, Jacobson, Jacqmin, Jang, Jani,
  Jaranowski, Jawahar, Jim\'enez-Forteza, Johnson, Johnson-McDaniel, Jones,
  Jones, Jonker, Ju, Haris, Kalaghatgi, Kalogera, Kandhasamy, Kang, Kanner,
  Karki, Kasprzack, Katsavounidis, Katzman, Kaufer, Kaur, Kawabe, Kawazoe,
  K\'ef\'elian, Kehl, Keitel, Kelley, Kells, Kennedy, Keppel, Key,
  Khalaidovski, Khalili, Khan, Khan, Khan, Khazanov, Kijbunchoo, Kim, Kim, Kim,
  Kim, Kim, Kim, King, King, Kinzel, Kissel, Kleybolte, Klimenko, Koehlenbeck,
  Kokeyama, Koley, Kondrashov, Kontos, Koranda, Korobko, Korth, Kowalska,
  Kozak, Kringel, Krishnan, Kr\'olak, Krueger, Kuehn, Kumar, Kumar, Kuo,
  Kutynia, Kwee, Lackey, Landry, Lange, Lantz, Lasky, Lazzarini, Lazzaro,
  Leaci, Leavey, Lebigot, Lee, Lee, Lee, Lee, Lenon, Leonardi, Leong, Leroy,
  Letendre, Levin, Levine, Li, Libson, Littenberg, Lockerbie, Logue, Lombardi,
  London, Lord, Lorenzini, Loriette, Lormand, Losurdo, Lough, Lousto, Lovelace,
  L\"uck, Lundgren, Luo, Lynch, Ma, MacDonald, Machenschalk, MacInnis, Macleod,
  Maga\~na Sandoval, Magee, Mageswaran, Majorana, Maksimovic, Malvezzi, Man,
  Mandel, Mandic, Mangano, Mansell, Manske, Mantovani, Marchesoni, Marion,
  M\'arka, M\'arka, Markosyan, Maros, Martelli, Martellini, Martin, Martin,
  Martynov, Marx, Mason, Masserot, Massinger, Masso-Reid, Matichard, Matone,
  Mavalvala, Mazumder, Mazzolo, McCarthy, McClelland, McCormick, McGuire,
  McIntyre, McIver, McManus, McWilliams, Meacher, Meadors, Meidam, Melatos,
  Mendell, Mendoza-Gandara, Mercer, Merilh, Merzougui, Meshkov, Messenger,
  Messick, Meyers, Mezzani, Miao, Michel, Middleton, Mikhailov, Milano, Miller,
  Millhouse, Minenkov, Ming, Mirshekari, Mishra, Mitra, Mitrofanov,
  Mitselmakher, Mittleman, Moggi, Mohan, Mohapatra, Montani, Moore, Moore,
  Moraru, Moreno, Morriss, Mossavi, Mours, Mow-Lowry, Mueller, Mueller, Muir,
  Mukherjee, Mukherjee, Mukherjee, Mukund, Mullavey, Munch, Murphy, Murray,
  Mytidis, Nardecchia, Naticchioni, Nayak, Necula, Nedkova, Nelemans, Neri,
  Neunzert, Newton, Nguyen, Nielsen, Nissanke, Nitz, Nocera, Nolting,
  Normandin, Nuttall, Oberling, Ochsner, O'Dell, Oelker, Ogin, Oh, Oh, Ohme,
  Oliver, Oppermann, Oram, O'Reilly, O'Shaughnessy, Ott, Ottaway, Ottens,
  Overmier, Owen, Pai, Pai, Palamos, Palashov, Palomba, Pal-Singh, Pan, Pan,
  Pankow, Pannarale, Pant, Paoletti, Paoli, Papa, Paris, Parker, Pascucci,
  Pasqualetti, Passaquieti, Passuello, Patricelli, Patrick, Pearlstone,
  Pedraza, Pedurand, Pekowsky, Pele, Penn, Perreca, Pfeiffer, Phelps, Piccinni,
  Pichot, Pickenpack, Piergiovanni, Pierro, Pillant, Pinard, Pinto, Pitkin,
  Poeld, Poggiani, Popolizio, Post, Powell, Prasad, Predoi, Premachandra,
  Prestegard, Price, Prijatelj, Principe, Privitera, Prix, Prodi, Prokhorov,
  Puncken, Punturo, Puppo, P\"urrer, Qi, Qin, Quetschke, Quintero,
  Quitzow-James, Raab, Rabeling, Radkins, Raffai, Raja, Rakhmanov, Ramet,
  Rapagnani, Raymond, Razzano, Re, Read, Reed, Regimbau, Rei, Reid, Reitze,
  Rew, Reyes, Ricci, Riles, Robertson, Robie, Robinet, Rocchi, Rolland,
  Rollins, Roma, Romano, Romano, Romanov, Romie, Rosi\ifmmode~\acute{n}\else
  \'{n}\fi{}ska, Rowan, R\"udiger, Ruggi, Ryan, Sachdev, Sadecki, Sadeghian,
  Salconi, Saleem, Salemi, Samajdar, Sammut, Sampson, Sanchez, Sandberg,
  Sandeen, Sanders, Sanders, Sassolas, Sathyaprakash, Saulson, Sauter, Savage,
  Sawadsky, Schale, Schilling, Schmidt, Schmidt, Schnabel, Schofield,
  Sch\"onbeck, Schreiber, Schuette, Schutz, Scott, Scott, Sellers, Sengupta,
  Sentenac, Sequino, Sergeev, Serna, Setyawati, Sevigny, Shaddock, Shaffer,
  Shah, Shahriar, Shaltev, Shao, Shapiro, Shawhan, Sheperd, Shoemaker,
  Shoemaker, Siellez, Siemens, Sigg, Silva, Simakov, Singer, Singer, Singh,
  Singh, Singhal, Sintes, Slagmolen, Smith, Smith, Smith, Smith, Son, Sorazu,
  Sorrentino, Souradeep, Srivastava, Staley, Steinke, Steinlechner,
  Steinlechner, Steinmeyer, Stephens, Stevenson, Stone, Strain, Straniero,
  Stratta, Strauss, Strigin, Sturani, Stuver, Summerscales, Sun, Sutton,
  Swinkels, Szczepa\ifmmode~\acute{n}\else \'{n}\fi{}czyk, Tacca, Talukder,
  Tanner, T\'apai, Tarabrin, Taracchini, Taylor, Theeg, Thirugnanasambandam,
  Thomas, Thomas, Thomas, Thorne, Thorne, Thrane, Tiwari, Tiwari, Tokmakov,
  Tomlinson, Tonelli, Torres, Torrie, T\"oyr\"a, Travasso, Traylor, Trifir\`o,
  Tringali, Trozzo, Tse, Turconi, Tuyenbayev, Ugolini, Unnikrishnan, Urban,
  Usman, Vahlbruch, Vajente, Valdes, Vallisneri, van Bakel, van Beuzekom,
  van~den Brand, Van Den~Broeck, Vander-Hyde, van~der Schaaf, van Heijningen,
  van Veggel, Vardaro, Vass, Vas\'uth, Vaulin, Vecchio, Vedovato, Veitch,
  Veitch, Venkateswara, Verkindt, Vetrano, Vicer\'e, Vinciguerra, Vine, Vinet,
  Vitale, Vo, Vocca, Vorvick, Voss, Vousden, Vyatchanin, Wade, Wade, Wade,
  Waldman, Walker, Wallace, Walsh, Wang, Wang, Wang, Wang, Wang, Ward, Ward,
  Warner, Was, Weaver, Wei, Weinert, Weinstein, Weiss, Welborn, Wen,
  We\ss{}els, Westphal, Wette, Whelan, Whitcomb, White, Whiting, Wiesner,
  Wilkinson, Willems, Williams, Williams, Williamson, Willis, Willke, Wimmer,
  Winkelmann, Winkler, Wipf, Wiseman, Wittel, Woan, Worden, Wright, Wu, Yablon,
  Yakushin, Yam, Yamamoto, Yancey, Yap, Yu, Yvert, Zadro\ifmmode~\dot{z}\else
  \.{z}\fi{}ny, Zangrando, Zanolin, Zendri, Zevin, Zhang, Zhang, Zhang, Zhang,
  Zhao, Zhou, Zhou, Zhu, Zucker, Zuraw, and Zweizig}{Abbott
  et~al.}{2016}]{ligo}
Abbott, B.~P., R.~Abbott, T.~D. Abbott, M.~R. Abernathy, F.~Acernese,
  K.~Ackley, C.~Adams, T.~Adams, P.~Addesso, R.~X. Adhikari, V.~B. Adya,
  C.~Affeldt, M.~Agathos, K.~Agatsuma, N.~Aggarwal, O.~D. Aguiar, L.~Aiello,
  A.~Ain, P.~Ajith, B.~Allen, A.~Allocca, P.~A. Altin, S.~B. Anderson, W.~G.
  Anderson, K.~Arai, M.~A. Arain, M.~C. Araya, C.~C. Arceneaux, J.~S. Areeda,
  N.~Arnaud, K.~G. Arun, S.~Ascenzi, G.~Ashton, M.~Ast, S.~M. Aston, P.~Astone,
  P.~Aufmuth, C.~Aulbert, S.~Babak, P.~Bacon, M.~K.~M. Bader, P.~T. Baker,
  F.~Baldaccini, G.~Ballardin, S.~W. Ballmer, J.~C. Barayoga, S.~E. Barclay,
  B.~C. Barish, D.~Barker, F.~Barone, B.~Barr, L.~Barsotti, M.~Barsuglia,
  D.~Barta, J.~Bartlett, M.~A. Barton, I.~Bartos, R.~Bassiri, A.~Basti, J.~C.
  Batch, C.~Baune, V.~Bavigadda, M.~Bazzan, B.~Behnke, M.~Bejger,
  C.~Belczynski, A.~S. Bell, C.~J. Bell, B.~K. Berger, J.~Bergman, G.~Bergmann,
  C.~P.~L. Berry, D.~Bersanetti, A.~Bertolini, J.~Betzwieser, S.~Bhagwat,
  R.~Bhandare, I.~A. Bilenko, G.~Billingsley, J.~Birch, R.~Birney,
  O.~Birnholtz, S.~Biscans, A.~Bisht, M.~Bitossi, C.~Biwer, M.~A. Bizouard,
  J.~K. Blackburn, C.~D. Blair, D.~G. Blair, R.~M. Blair, S.~Bloemen, O.~Bock,
  T.~P. Bodiya, M.~Boer, G.~Bogaert, C.~Bogan, A.~Bohe, P.~Bojtos, C.~Bond,
  F.~Bondu, R.~Bonnand, B.~A. Boom, R.~Bork, V.~Boschi, S.~Bose, Y.~Bouffanais,
  A.~Bozzi, C.~Bradaschia, P.~R. Brady, V.~B. Braginsky, M.~Branchesi, J.~E.
  Brau, T.~Briant, A.~Brillet, M.~Brinkmann, V.~Brisson, P.~Brockill, A.~F.
  Brooks, D.~A. Brown, D.~D. Brown, N.~M. Brown, C.~C. Buchanan, A.~Buikema,
  T.~Bulik, H.~J. Bulten, A.~Buonanno, D.~Buskulic, C.~Buy, R.~L. Byer,
  M.~Cabero, L.~Cadonati, G.~Cagnoli, C.~Cahillane, J.~C. Bustillo,
  T.~Callister, E.~Calloni, J.~B. Camp, K.~C. Cannon, J.~Cao, C.~D. Capano,
  E.~Capocasa, F.~Carbognani, S.~Caride, J.~C. Diaz, C.~Casentini, S.~Caudill,
  M.~Cavagli\`a, F.~Cavalier, R.~Cavalieri, G.~Cella, C.~B. Cepeda, L.~C.
  Baiardi, G.~Cerretani, E.~Cesarini, R.~Chakraborty, T.~Chalermsongsak, S.~J.
  Chamberlin, M.~Chan, S.~Chao, P.~Charlton, E.~Chassande-Mottin, H.~Y. Chen,
  Y.~Chen, C.~Cheng, A.~Chincarini, A.~Chiummo, H.~S. Cho, M.~Cho, J.~H. Chow,
  N.~Christensen, Q.~Chu, S.~Chua, S.~Chung, G.~Ciani, F.~Clara, J.~A. Clark,
  F.~Cleva, E.~Coccia, P.-F. Cohadon, A.~Colla, C.~G. Collette, L.~Cominsky,
  M.~Constancio, A.~Conte, L.~Conti, D.~Cook, T.~R. Corbitt, N.~Cornish,
  A.~Corsi, S.~Cortese, C.~A. Costa, M.~W. Coughlin, S.~B. Coughlin, J.-P.
  Coulon, S.~T. Countryman, P.~Couvares, E.~E. Cowan, D.~M. Coward, M.~J.
  Cowart, D.~C. Coyne, R.~Coyne, K.~Craig, J.~D.~E. Creighton, T.~D. Creighton,
  J.~Cripe, S.~G. Crowder, A.~M. Cruise, A.~Cumming, L.~Cunningham, E.~Cuoco,
  T.~D. Canton, S.~L. Danilishin, S.~D'Antonio, K.~Danzmann, N.~S. Darman,
  C.~F. Da~Silva~Costa, V.~Dattilo, I.~Dave, H.~P. Daveloza, M.~Davier, G.~S.
  Davies, E.~J. Daw, R.~Day, S.~De, D.~DeBra, G.~Debreczeni, J.~Degallaix,
  M.~De~Laurentis, S.~Del\'eglise, W.~Del~Pozzo, T.~Denker, T.~Dent, H.~Dereli,
  V.~Dergachev, R.~T. DeRosa, R.~De~Rosa, R.~DeSalvo, S.~Dhurandhar, M.~C.
  D\'{\i}az, L.~Di~Fiore, M.~Di~Giovanni, A.~Di~Lieto, S.~Di~Pace, I.~Di~Palma,
  A.~Di~Virgilio, G.~Dojcinoski, V.~Dolique, F.~Donovan, K.~L. Dooley,
  S.~Doravari, R.~Douglas, T.~P. Downes, M.~Drago, R.~W.~P. Drever, J.~C.
  Driggers, Z.~Du, M.~Ducrot, S.~E. Dwyer, T.~B. Edo, M.~C. Edwards, A.~Effler,
  H.-B. Eggenstein, P.~Ehrens, J.~Eichholz, S.~S. Eikenberry, W.~Engels, R.~C.
  Essick, T.~Etzel, M.~Evans, T.~M. Evans, R.~Everett, M.~Factourovich,
  V.~Fafone, H.~Fair, S.~Fairhurst, X.~Fan, Q.~Fang, S.~Farinon, B.~Farr, W.~M.
  Farr, M.~Favata, M.~Fays, H.~Fehrmann, M.~M. Fejer, D.~Feldbaum, I.~Ferrante,
  E.~C. Ferreira, F.~Ferrini, F.~Fidecaro, L.~S. Finn, I.~Fiori, D.~Fiorucci,
  R.~P. Fisher, R.~Flaminio, M.~Fletcher, H.~Fong, J.-D. Fournier, S.~Franco,
  S.~Frasca, F.~Frasconi, M.~Frede, Z.~Frei, A.~Freise, R.~Frey, V.~Frey, T.~T.
  Fricke, P.~Fritschel, V.~V. Frolov, P.~Fulda, M.~Fyffe, H.~A.~G. Gabbard,
  J.~R. Gair, L.~Gammaitoni, S.~G. Gaonkar, F.~Garufi, A.~Gatto, G.~Gaur,
  N.~Gehrels, G.~Gemme, B.~Gendre, E.~Genin, A.~Gennai, J.~George, L.~Gergely,
  V.~Germain, A.~Ghosh, A.~Ghosh, S.~Ghosh, J.~A. Giaime, K.~D. Giardina,
  A.~Giazotto, K.~Gill, A.~Glaefke, J.~R. Gleason, E.~Goetz, R.~Goetz,
  L.~Gondan, G.~Gonz\'alez, J.~M.~G. Castro, A.~Gopakumar, N.~A. Gordon, M.~L.
  Gorodetsky, S.~E. Gossan, M.~Gosselin, R.~Gouaty, C.~Graef, P.~B. Graff,
  M.~Granata, A.~Grant, S.~Gras, C.~Gray, G.~Greco, A.~C. Green, R.~J.~S.
  Greenhalgh, P.~Groot, H.~Grote, S.~Grunewald, G.~M. Guidi, X.~Guo, A.~Gupta,
  M.~K. Gupta, K.~E. Gushwa, E.~K. Gustafson, R.~Gustafson, J.~J. Hacker, B.~R.
  Hall, E.~D. Hall, G.~Hammond, M.~Haney, M.~M. Hanke, J.~Hanks, C.~Hanna,
  M.~D. Hannam, J.~Hanson, T.~Hardwick, J.~Harms, G.~M. Harry, I.~W. Harry,
  M.~J. Hart, M.~T. Hartman, C.-J. Haster, K.~Haughian, J.~Healy, J.~Heefner,
  A.~Heidmann, M.~C. Heintze, G.~Heinzel, H.~Heitmann, P.~Hello, G.~Hemming,
  M.~Hendry, I.~S. Heng, J.~Hennig, A.~W. Heptonstall, M.~Heurs, S.~Hild,
  D.~Hoak, K.~A. Hodge, D.~Hofman, S.~E. Hollitt, K.~Holt, D.~E. Holz,
  P.~Hopkins, D.~J. Hosken, J.~Hough, E.~A. Houston, E.~J. Howell, Y.~M. Hu,
  S.~Huang, E.~A. Huerta, D.~Huet, B.~Hughey, S.~Husa, S.~H. Huttner,
  T.~Huynh-Dinh, A.~Idrisy, N.~Indik, D.~R. Ingram, R.~Inta, H.~N. Isa, J.-M.
  Isac, M.~Isi, G.~Islas, T.~Isogai, B.~R. Iyer, K.~Izumi, M.~B. Jacobson,
  T.~Jacqmin, H.~Jang, K.~Jani, P.~Jaranowski, S.~Jawahar,
  F.~Jim\'enez-Forteza, W.~W. Johnson, N.~K. Johnson-McDaniel, D.~I. Jones,
  R.~Jones, R.~J.~G. Jonker, L.~Ju, K.~Haris, C.~V. Kalaghatgi, V.~Kalogera,
  S.~Kandhasamy, G.~Kang, J.~B. Kanner, S.~Karki, M.~Kasprzack,
  E.~Katsavounidis, W.~Katzman, S.~Kaufer, T.~Kaur, K.~Kawabe, F.~Kawazoe,
  F.~K\'ef\'elian, M.~S. Kehl, D.~Keitel, D.~B. Kelley, W.~Kells, R.~Kennedy,
  D.~G. Keppel, J.~S. Key, A.~Khalaidovski, F.~Y. Khalili, I.~Khan, S.~Khan,
  Z.~Khan, E.~A. Khazanov, N.~Kijbunchoo, C.~Kim, J.~Kim, K.~Kim, N.-G. Kim,
  N.~Kim, Y.-M. Kim, E.~J. King, P.~J. King, D.~L. Kinzel, J.~S. Kissel,
  L.~Kleybolte, S.~Klimenko, S.~M. Koehlenbeck, K.~Kokeyama, S.~Koley,
  V.~Kondrashov, A.~Kontos, S.~Koranda, M.~Korobko, W.~Z. Korth, I.~Kowalska,
  D.~B. Kozak, V.~Kringel, B.~Krishnan, A.~Kr\'olak, C.~Krueger, G.~Kuehn,
  P.~Kumar, R.~Kumar, L.~Kuo, A.~Kutynia, P.~Kwee, B.~D. Lackey, M.~Landry,
  J.~Lange, B.~Lantz, P.~D. Lasky, A.~Lazzarini, C.~Lazzaro, P.~Leaci,
  S.~Leavey, E.~O. Lebigot, C.~H. Lee, H.~K. Lee, H.~M. Lee, K.~Lee, A.~Lenon,
  M.~Leonardi, J.~R. Leong, N.~Leroy, N.~Letendre, Y.~Levin, B.~M. Levine,
  T.~G.~F. Li, A.~Libson, T.~B. Littenberg, N.~A. Lockerbie, J.~Logue, A.~L.
  Lombardi, L.~T. London, J.~E. Lord, M.~Lorenzini, V.~Loriette, M.~Lormand,
  G.~Losurdo, J.~D. Lough, C.~O. Lousto, G.~Lovelace, H.~L\"uck, A.~P.
  Lundgren, J.~Luo, R.~Lynch, Y.~Ma, T.~MacDonald, B.~Machenschalk,
  M.~MacInnis, D.~M. Macleod, F.~Maga\~na Sandoval, R.~M. Magee, M.~Mageswaran,
  E.~Majorana, I.~Maksimovic, V.~Malvezzi, N.~Man, I.~Mandel, V.~Mandic,
  V.~Mangano, G.~L. Mansell, M.~Manske, M.~Mantovani, F.~Marchesoni, F.~Marion,
  S.~M\'arka, Z.~M\'arka, A.~S. Markosyan, E.~Maros, F.~Martelli,
  L.~Martellini, I.~W. Martin, R.~M. Martin, D.~V. Martynov, J.~N. Marx,
  K.~Mason, A.~Masserot, T.~J. Massinger, M.~Masso-Reid, F.~Matichard,
  L.~Matone, N.~Mavalvala, N.~Mazumder, G.~Mazzolo, R.~McCarthy, D.~E.
  McClelland, S.~McCormick, S.~C. McGuire, G.~McIntyre, J.~McIver, D.~J.
  McManus, S.~T. McWilliams, D.~Meacher, G.~D. Meadors, J.~Meidam, A.~Melatos,
  G.~Mendell, D.~Mendoza-Gandara, R.~A. Mercer, E.~Merilh, M.~Merzougui,
  S.~Meshkov, C.~Messenger, C.~Messick, P.~M. Meyers, F.~Mezzani, H.~Miao,
  C.~Michel, H.~Middleton, E.~E. Mikhailov, L.~Milano, J.~Miller, M.~Millhouse,
  Y.~Minenkov, J.~Ming, S.~Mirshekari, C.~Mishra, S.~Mitra, V.~P. Mitrofanov,
  G.~Mitselmakher, R.~Mittleman, A.~Moggi, M.~Mohan, S.~R.~P. Mohapatra,
  M.~Montani, B.~C. Moore, C.~J. Moore, D.~Moraru, G.~Moreno, S.~R. Morriss,
  K.~Mossavi, B.~Mours, C.~M. Mow-Lowry, C.~L. Mueller, G.~Mueller, A.~W. Muir,
  A.~Mukherjee, D.~Mukherjee, S.~Mukherjee, N.~Mukund, A.~Mullavey, J.~Munch,
  D.~J. Murphy, P.~G. Murray, A.~Mytidis, I.~Nardecchia, L.~Naticchioni, R.~K.
  Nayak, V.~Necula, K.~Nedkova, G.~Nelemans, M.~Neri, A.~Neunzert, G.~Newton,
  T.~T. Nguyen, A.~B. Nielsen, S.~Nissanke, A.~Nitz, F.~Nocera, D.~Nolting,
  M.~E.~N. Normandin, L.~K. Nuttall, J.~Oberling, E.~Ochsner, J.~O'Dell,
  E.~Oelker, G.~H. Ogin, J.~J. Oh, S.~H. Oh, F.~Ohme, M.~Oliver, P.~Oppermann,
  R.~J. Oram, B.~O'Reilly, R.~O'Shaughnessy, C.~D. Ott, D.~J. Ottaway, R.~S.
  Ottens, H.~Overmier, B.~J. Owen, A.~Pai, S.~A. Pai, J.~R. Palamos,
  O.~Palashov, C.~Palomba, A.~Pal-Singh, H.~Pan, Y.~Pan, C.~Pankow,
  F.~Pannarale, B.~C. Pant, F.~Paoletti, A.~Paoli, M.~A. Papa, H.~R. Paris,
  W.~Parker, D.~Pascucci, A.~Pasqualetti, R.~Passaquieti, D.~Passuello,
  B.~Patricelli, Z.~Patrick, B.~L. Pearlstone, M.~Pedraza, R.~Pedurand,
  L.~Pekowsky, A.~Pele, S.~Penn, A.~Perreca, H.~P. Pfeiffer, M.~Phelps,
  O.~Piccinni, M.~Pichot, M.~Pickenpack, F.~Piergiovanni, V.~Pierro,
  G.~Pillant, L.~Pinard, I.~M. Pinto, M.~Pitkin, J.~H. Poeld, R.~Poggiani,
  P.~Popolizio, A.~Post, J.~Powell, J.~Prasad, V.~Predoi, S.~S. Premachandra,
  T.~Prestegard, L.~R. Price, M.~Prijatelj, M.~Principe, S.~Privitera, R.~Prix,
  G.~A. Prodi, L.~Prokhorov, O.~Puncken, M.~Punturo, P.~Puppo, M.~P\"urrer,
  H.~Qi, J.~Qin, V.~Quetschke, E.~A. Quintero, R.~Quitzow-James, F.~J. Raab,
  D.~S. Rabeling, H.~Radkins, P.~Raffai, S.~Raja, M.~Rakhmanov, C.~R. Ramet,
  P.~Rapagnani, V.~Raymond, M.~Razzano, V.~Re, J.~Read, C.~M. Reed,
  T.~Regimbau, L.~Rei, S.~Reid, D.~H. Reitze, H.~Rew, S.~D. Reyes, F.~Ricci,
  K.~Riles, N.~A. Robertson, R.~Robie, F.~Robinet, A.~Rocchi, L.~Rolland, J.~G.
  Rollins, V.~J. Roma, J.~D. Romano, R.~Romano, G.~Romanov, J.~H. Romie,
  D.~Rosi\ifmmode~\acute{n}\else \'{n}\fi{}ska, S.~Rowan, A.~R\"udiger,
  P.~Ruggi, K.~Ryan, S.~Sachdev, T.~Sadecki, L.~Sadeghian, L.~Salconi,
  M.~Saleem, F.~Salemi, A.~Samajdar, L.~Sammut, L.~M. Sampson, E.~J. Sanchez,
  V.~Sandberg, B.~Sandeen, G.~H. Sanders, J.~R. Sanders, B.~Sassolas, B.~S.
  Sathyaprakash, P.~R. Saulson, O.~Sauter, R.~L. Savage, A.~Sawadsky,
  P.~Schale, R.~Schilling, J.~Schmidt, P.~Schmidt, R.~Schnabel, R.~M.~S.
  Schofield, A.~Sch\"onbeck, E.~Schreiber, D.~Schuette, B.~F. Schutz, J.~Scott,
  S.~M. Scott, D.~Sellers, A.~S. Sengupta, D.~Sentenac, V.~Sequino, A.~Sergeev,
  G.~Serna, Y.~Setyawati, A.~Sevigny, D.~A. Shaddock, T.~Shaffer, S.~Shah,
  M.~S. Shahriar, M.~Shaltev, Z.~Shao, B.~Shapiro, P.~Shawhan, A.~Sheperd,
  D.~H. Shoemaker, D.~M. Shoemaker, K.~Siellez, X.~Siemens, D.~Sigg, A.~D.
  Silva, D.~Simakov, A.~Singer, L.~P. Singer, A.~Singh, R.~Singh, A.~Singhal,
  A.~M. Sintes, B.~J.~J. Slagmolen, J.~R. Smith, M.~R. Smith, N.~D. Smith,
  R.~J.~E. Smith, E.~J. Son, B.~Sorazu, F.~Sorrentino, T.~Souradeep, A.~K.
  Srivastava, A.~Staley, M.~Steinke, J.~Steinlechner, S.~Steinlechner,
  D.~Steinmeyer, B.~C. Stephens, S.~P. Stevenson, R.~Stone, K.~A. Strain,
  N.~Straniero, G.~Stratta, N.~A. Strauss, S.~Strigin, R.~Sturani, A.~L.
  Stuver, T.~Z. Summerscales, L.~Sun, P.~J. Sutton, B.~L. Swinkels, M.~J.
  Szczepa\ifmmode~\acute{n}\else \'{n}\fi{}czyk, M.~Tacca, D.~Talukder, D.~B.
  Tanner, M.~T\'apai, S.~P. Tarabrin, A.~Taracchini, R.~Taylor, T.~Theeg, M.~P.
  Thirugnanasambandam, E.~G. Thomas, M.~Thomas, P.~Thomas, K.~A. Thorne, K.~S.
  Thorne, E.~Thrane, S.~Tiwari, V.~Tiwari, K.~V. Tokmakov, C.~Tomlinson,
  M.~Tonelli, C.~V. Torres, C.~I. Torrie, D.~T\"oyr\"a, F.~Travasso,
  G.~Traylor, D.~Trifir\`o, M.~C. Tringali, L.~Trozzo, M.~Tse, M.~Turconi,
  D.~Tuyenbayev, D.~Ugolini, C.~S. Unnikrishnan, A.~L. Urban, S.~A. Usman,
  H.~Vahlbruch, G.~Vajente, G.~Valdes, M.~Vallisneri, N.~van Bakel, M.~van
  Beuzekom, J.~F.~J. van~den Brand, C.~Van Den~Broeck, D.~C. Vander-Hyde,
  L.~van~der Schaaf, J.~V. van Heijningen, A.~A. van Veggel, M.~Vardaro,
  S.~Vass, M.~Vas\'uth, R.~Vaulin, A.~Vecchio, G.~Vedovato, J.~Veitch, P.~J.
  Veitch, K.~Venkateswara, D.~Verkindt, F.~Vetrano, A.~Vicer\'e,
  S.~Vinciguerra, D.~J. Vine, J.-Y. Vinet, S.~Vitale, T.~Vo, H.~Vocca,
  C.~Vorvick, D.~Voss, W.~D. Vousden, S.~P. Vyatchanin, A.~R. Wade, L.~E. Wade,
  M.~Wade, S.~J. Waldman, M.~Walker, L.~Wallace, S.~Walsh, G.~Wang, H.~Wang,
  M.~Wang, X.~Wang, Y.~Wang, H.~Ward, R.~L. Ward, J.~Warner, M.~Was, B.~Weaver,
  L.-W. Wei, M.~Weinert, A.~J. Weinstein, R.~Weiss, T.~Welborn, L.~Wen,
  P.~We\ss{}els, T.~Westphal, K.~Wette, J.~T. Whelan, S.~E. Whitcomb, D.~J.
  White, B.~F. Whiting, K.~Wiesner, C.~Wilkinson, P.~A. Willems, L.~Williams,
  R.~D. Williams, A.~R. Williamson, J.~L. Willis, B.~Willke, M.~H. Wimmer,
  L.~Winkelmann, W.~Winkler, C.~C. Wipf, A.~G. Wiseman, H.~Wittel, G.~Woan,
  J.~Worden, J.~L. Wright, G.~Wu, J.~Yablon, I.~Yakushin, W.~Yam, H.~Yamamoto,
  C.~C. Yancey, M.~J. Yap, H.~Yu, M.~Yvert, A.~Zadro\ifmmode~\dot{z}\else
  \.{z}\fi{}ny, L.~Zangrando, M.~Zanolin, J.-P. Zendri, M.~Zevin, F.~Zhang,
  L.~Zhang, M.~Zhang, Y.~Zhang, C.~Zhao, M.~Zhou, Z.~Zhou, X.~J. Zhu, M.~E.
  Zucker, S.~E. Zuraw, and J.~Zweizig (2016, Feb).
\newblock Observation of gravitational waves from a binary black hole merger.
\newblock {\em Phys. Rev. Lett.\/}~{\em 116}, 061102.

\bibitem[\protect\citeauthoryear{Arrilucea, Commission, for Research, and
  Innovation}{Arrilucea et~al.}{2018}]{apollo}
Arrilucea, E., E.~Commission, D.-G. for Research, and Innovation (2018).
\newblock {\em Mission-oriented R\&I policies : in-depth case studies : Apollo
  project (US) : case study report}.
\newblock Publications Office.

\bibitem[\protect\citeauthoryear{Beaulieu and Leonelli}{Beaulieu and
  Leonelli}{2021}]{beaulieu-leonelli-2021-data}
Beaulieu, A. and S.~Leonelli (2021).
\newblock {\em Data and Society: A Critical Introduction}.
\newblock Sage.

\bibitem[\protect\citeauthoryear{Bekman}{Bekman}{2022}]{bekman-2022-bloom}
Bekman, S. (2022, Jul).
\newblock The technology behind bloom training.
\newblock {\em Hugging Face Blog\/}.

\bibitem[\protect\citeauthoryear{Bekman and Gugger}{Bekman and
  Gugger}{2022}]{bekman-2022-accelerate}
Bekman, S. and S.~Gugger (2022, Sep).
\newblock Incredibly fast bloom inference with deepspeed and accelerate.
\newblock {\em Hugging Face Blog\/}.

\bibitem[\protect\citeauthoryear{Bender, Gebru, McMillan-Major, and
  Shmitchell}{Bender et~al.}{2021}]{stochastic}
Bender, E.~M., T.~Gebru, A.~McMillan-Major, and S.~Shmitchell (2021).
\newblock On the dangers of stochastic parrots: Can language models be too big?
\newblock In {\em Proceedings of the 2021 ACM Conference on Fairness,
  Accountability, and Transparency}, FAccT '21, New York, NY, USA, pp.\
  610–623. Association for Computing Machinery.

\bibitem[\protect\citeauthoryear{Birhane, Isaac, Prabhakaran, Díaz, Elish,
  Gabriel, and Mohamed}{Birhane et~al.}{2022}]{birhane-participatory}
Birhane, A., W.~Isaac, V.~Prabhakaran, M.~Díaz, M.~Elish, I.~Gabriel, and
  S.~Mohamed (2022).
\newblock Power to the people? opportunities and challenges for participatory
  ai.
\newblock In {\em Equity and Access in Algorithms, Mechanisms, and
  Optimization}, EAAMO '22, New York, NY, USA. Association for Computing
  Machinery.

\bibitem[\protect\citeauthoryear{Birhane, Kalluri, Card, Agnew, Dotan, and
  Bao}{Birhane et~al.}{2022}]{values-ml}
Birhane, A., P.~Kalluri, D.~Card, W.~Agnew, R.~Dotan, and M.~Bao (2022).
\newblock The values encoded in machine learning research.
\newblock In {\em 2022 ACM Conference on Fairness, Accountability, and
  Transparency}, FAccT '22, New York, NY, USA, pp.\  173–184. Association for
  Computing Machinery.

\bibitem[\protect\citeauthoryear{Birhane, Prabhu, and Kahembwe}{Birhane
  et~al.}{2021a}]{birhane-laion}
Birhane, A., V.~U. Prabhu, and E.~Kahembwe (2021a).
\newblock Multimodal datasets: misogyny, pornography, and malignant
  stereotypes.
\newblock {\em CoRR\/}~{\em abs/2110.01963}.

\bibitem[\protect\citeauthoryear{Birhane, Prabhu, and Kahembwe}{Birhane
  et~al.}{2021b}]{laion-analysis}
Birhane, A., V.~U. Prabhu, and E.~Kahembwe (2021b).
\newblock Multimodal datasets: misogyny, pornography, and malignant
  stereotypes.
\newblock {\em ArXiv\/}~{\em abs/2110.01963}.

\bibitem[\protect\citeauthoryear{Brown, Mann, Ryder, Subbiah, Kaplan, Dhariwal,
  Neelakantan, Shyam, Sastry, Askell, Agarwal, Herbert-Voss, Krueger, Henighan,
  Child, Ramesh, Ziegler, Wu, Winter, Hesse, Chen, Sigler, Litwin, Gray, Chess,
  Clark, Berner, McCandlish, Radford, Sutskever, and Amodei}{Brown
  et~al.}{2020}]{gpt3}
Brown, T., B.~Mann, N.~Ryder, M.~Subbiah, J.~D. Kaplan, P.~Dhariwal,
  A.~Neelakantan, P.~Shyam, G.~Sastry, A.~Askell, S.~Agarwal, A.~Herbert-Voss,
  G.~Krueger, T.~Henighan, R.~Child, A.~Ramesh, D.~Ziegler, J.~Wu, C.~Winter,
  C.~Hesse, M.~Chen, E.~Sigler, M.~Litwin, S.~Gray, B.~Chess, J.~Clark,
  C.~Berner, S.~McCandlish, A.~Radford, I.~Sutskever, and D.~Amodei (2020).
\newblock Language models are few-shot learners.
\newblock In H.~Larochelle, M.~Ranzato, R.~Hadsell, M.~Balcan, and H.~Lin
  (Eds.), {\em Advances in Neural Information Processing Systems}, Volume~33,
  pp.\  1877--1901. Curran Associates, Inc.

\bibitem[\protect\citeauthoryear{Community}{Community}{2022}]{the_turing_way_community_2022_6909298}
Community, T. T.~W. (2022, July).
\newblock {The Turing Way: A handbook for reproducible, ethical and
  collaborative research}.

\bibitem[\protect\citeauthoryear{Dodge, Marasovi{\'c}, Ilharco, Groeneveld,
  Mitchell, and Gardner}{Dodge et~al.}{2021}]{c4-analysis}
Dodge, J., A.~Marasovi{\'c}, G.~Ilharco, D.~Groeneveld, M.~Mitchell, and
  M.~Gardner (2021).
\newblock Documenting large webtext corpora: A case study on the colossal clean
  crawled corpus.
\newblock In {\em EMNLP}.

\bibitem[\protect\citeauthoryear{Elliott}{Elliott}{2017}]{elliott-tapestry-values}
Elliott, K.~C. (2017, 02).
\newblock {\em {A Tapestry of Values: An Introduction to Values in Science}}.
\newblock Oxford University Press.

\bibitem[\protect\citeauthoryear{Fan, Ilic, Wolf, and Gall{\'e}}{Fan
  et~al.}{2022}]{bigscience-2022-bigscience}
Fan, A., S.~Ilic, T.~Wolf, and M.~Gall{\'e} (Eds.) (2022, May).
\newblock {\em Proceedings of BigScience Episode {\#}5 -- Workshop on
  Challenges {\&} Perspectives in Creating Large Language Models},
  virtual+Dublin. Association for Computational Linguistics.

\bibitem[\protect\citeauthoryear{Gao, Tow, Biderman, Black, DiPofi, Foster,
  Golding, Hsu, McDonell, Muennighoff, Phang, Reynolds, Tang, Thite, Wang,
  Wang, and Zou}{Gao et~al.}{2021}]{eval-harness}
Gao, L., J.~Tow, S.~Biderman, S.~Black, A.~DiPofi, C.~Foster, L.~Golding,
  J.~Hsu, K.~McDonell, N.~Muennighoff, J.~Phang, L.~Reynolds, E.~Tang,
  A.~Thite, B.~Wang, K.~Wang, and A.~Zou (2021, September).
\newblock A framework for few-shot language model evaluation.

\bibitem[\protect\citeauthoryear{Haelewaters, Hofmann, and
  Romero-Olivares}{Haelewaters et~al.}{2021}]{helicopter-research}
Haelewaters, D., T.~A. Hofmann, and A.~L. Romero-Olivares (2021).
\newblock Ten simple rules for global north researchers to stop perpetuating
  helicopter research in the global south.
\newblock {\em PLoS Computational Biology\/}~{\em 17}.

\bibitem[\protect\citeauthoryear{Haidt}{Haidt}{2002}]{moral-emotions}
Haidt, J. (2002).
\newblock The moral emotions.
\newblock In R.~Davidson, K.~Scherer, and H.~Goldsmith (Eds.), {\em Handbook of
  Affective Sciences}, Series in affective science, Chapter~45. Oxford
  University Press.

\bibitem[\protect\citeauthoryear{Heathwood}{Heathwood}{2015}]{handbook-value}
Heathwood, C. (2015).
\newblock Monism and pluralism about value.
\newblock In I.~Hirose and J.~Olson (Eds.), {\em The Oxford Handbook of Value
  Theory}, Chapter~8, pp.\  136--155. Oxford University Press.

\bibitem[\protect\citeauthoryear{Hey, Tansley, Tolle, and Gray}{Hey
  et~al.}{2009}]{hey2009-fourth-paradigm}
Hey, T., S.~Tansley, K.~Tolle, and J.~Gray (2009, October).
\newblock {\em The Fourth Paradigm: Data-Intensive Scientific Discovery}.
\newblock Microsoft Research.

\bibitem[\protect\citeauthoryear{Hoffmann}{Hoffmann}{2021}]{terms-of-inclusion}
Hoffmann, A.~L. (2021).
\newblock Terms of inclusion: Data, discourse, violence.
\newblock {\em New Media \& Society\/}~{\em 23}, 3539 -- 3556.

\bibitem[\protect\citeauthoryear{Joshi, Santy, Budhiraja, Bali, and
  Choudhury}{Joshi et~al.}{2020}]{nlp-ling-diversity}
Joshi, P.~M., S.~Santy, A.~Budhiraja, K.~Bali, and M.~Choudhury (2020).
\newblock The state and fate of linguistic diversity and inclusion in the nlp
  world.
\newblock In {\em ACL}.

\bibitem[\protect\citeauthoryear{Krohs}{Krohs}{2012}]{krohs-convenience-experimentation}
Krohs, U. (2012).
\newblock Convenience experimentation.
\newblock {\em Studies in History and Philosophy of Science Part C: Studies in
  History and Philosophy of Biological and Biomedical Sciences\/}~{\em
  43\/}(1), 52--57.
\newblock Data-Driven Research in the Biological and Biomedical Sciences On
  Nature and Normativity: Normativity, Teleology, and Mechanism in Biological
  Explanation.

\bibitem[\protect\citeauthoryear{Kuhn}{Kuhn}{1962}]{kuhn:1962}
Kuhn, T.~S. (1962).
\newblock {\em {The Structure of Scientific Revolutions}}.
\newblock Chicago: University of Chicago Press.

\bibitem[\protect\citeauthoryear{Lander, Linton, Birren, Nusbaum, Zody,
  Baldwin, Devon, Dewar, Doyle, Fitzhugh, Funke, Gaige, Harris, Heaford,
  Howland, Kann, Lehoczky, LeVine, McEwan, and Koculi}{Lander
  et~al.}{2001}]{human-genome}
Lander, E., L.~Linton, B.~Birren, C.~Nusbaum, M.~Zody, J.~Baldwin, K.~Devon,
  K.~Dewar, M.~Doyle, W.~Fitzhugh, R.~Funke, D.~Gaige, K.~Harris, A.~Heaford,
  J.~Howland, L.~Kann, J.~Lehoczky, R.~LeVine, P.~McEwan, and E.~Koculi (2001,
  03).
\newblock Initial sequencing and analysis of the human genome.
\newblock {\em Nature\/}~{\em 409}, 860--921.

\bibitem[\protect\citeauthoryear{Lauren{\c{c}}on, Saulnier, Wang, Akiki, del
  Moral, Scao, Werra, Mou, Ponferrada, Nguyen, Frohberg, {\v{S}}a{\v{s}}ko,
  Lhoest, McMillan-Major, Dupont, Biderman, Rogers, allal, Toni, Pistilli,
  Nguyen, Nikpoor, Masoud, Colombo, de~la Rosa, Villegas, Thrush, Longpre,
  Nagel, Weber, Mu{\~n}oz, Zhu, Strien, Alyafeai, Almubarak, Chien,
  Gonzalez-Dios, Soroa, Lo, Dey, Suarez, Gokaslan, Bose, Adelani, Phan, Tran,
  Yu, Pai, Chim, Lepercq, Ilic, Mitchell, Luccioni, and
  Jernite}{Lauren{\c{c}}on et~al.}{2022}]{roots}
Lauren{\c{c}}on, H., L.~Saulnier, T.~Wang, C.~Akiki, A.~V. del Moral, T.~L.
  Scao, L.~V. Werra, C.~Mou, E.~G. Ponferrada, H.~Nguyen, J.~Frohberg,
  M.~{\v{S}}a{\v{s}}ko, Q.~Lhoest, A.~McMillan-Major, G.~Dupont, S.~Biderman,
  A.~Rogers, L.~B. allal, F.~D. Toni, G.~Pistilli, O.~Nguyen, S.~Nikpoor,
  M.~Masoud, P.~Colombo, J.~de~la Rosa, P.~Villegas, T.~Thrush, S.~Longpre,
  S.~Nagel, L.~Weber, M.~R. Mu{\~n}oz, J.~Zhu, D.~V. Strien, Z.~Alyafeai,
  K.~Almubarak, V.~M. Chien, I.~Gonzalez-Dios, A.~Soroa, K.~Lo, M.~Dey, P.~O.
  Suarez, A.~Gokaslan, S.~Bose, D.~I. Adelani, L.~Phan, H.~Tran, I.~Yu, S.~Pai,
  J.~Chim, V.~Lepercq, S.~Ilic, M.~Mitchell, S.~Luccioni, and Y.~Jernite
  (2022).
\newblock The bigscience {ROOTS} corpus: A 1.6{TB} composite multilingual
  dataset.
\newblock In {\em Thirty-sixth Conference on Neural Information Processing
  Systems Datasets and Benchmarks Track}.

\bibitem[\protect\citeauthoryear{Leonelli}{Leonelli}{2020}]{sep-leonelli-science-big-data}
Leonelli, S. (2020).
\newblock {Scientific Research and Big Data}.
\newblock In E.~N. Zalta (Ed.), {\em The {Stanford} Encyclopedia of
  Philosophy\/} ({S}ummer 2020 ed.). Metaphysics Research Lab, Stanford
  University.

\bibitem[\protect\citeauthoryear{Li}{Li}{2006}]{harmony}
Li, C. (2006).
\newblock The confucian ideal of harmony.
\newblock {\em Philosophy East and West\/}~{\em 56\/}(4), 583--603.

\bibitem[\protect\citeauthoryear{Longino}{Longino}{2019}]{sep-longino-scientific-knowledge-social}
Longino, H. (2019).
\newblock {The Social Dimensions of Scientific Knowledge}.
\newblock In E.~N. Zalta (Ed.), {\em The {Stanford} Encyclopedia of
  Philosophy\/} ({S}ummer 2019 ed.). Metaphysics Research Lab, Stanford
  University.

\bibitem[\protect\citeauthoryear{McMillan-Major, Alyafeai, Biderman, Chen,
  Toni, Dupont, ElSahar, Emezue, Aji, Ili'c, Khamis, Leong, Masoud, Etxabe,
  Suarez, Talat, van Strien, and Jernite}{McMillan-Major
  et~al.}{2022}]{bigscience-catalog}
McMillan-Major, A., Z.~Alyafeai, S.~R. Biderman, K.~Chen, F.~D. Toni,
  G.~Dupont, H.~ElSahar, C.~C. Emezue, A.~F. Aji, S.~Ili'c, N.~Khamis,
  C.~Leong, M.~Masoud, A.~S. Etxabe, P.~O. Suarez, Z.~Talat, D.~A. van Strien,
  and Y.~Jernite (2022).
\newblock Documenting geographically and contextually diverse data sources: The
  bigscience catalogue of language data and resources.
\newblock {\em ArXiv\/}~{\em abs/2201.10066}.

\bibitem[\protect\citeauthoryear{Park, Lee, Jang, and Jung}{Park
  et~al.}{2020}]{korean-tokenization}
Park, K., J.~Lee, S.~Jang, and D.~Jung (2020).
\newblock An empirical study of tokenization strategies for various korean nlp
  tasks.
\newblock In {\em AACL}.

\bibitem[\protect\citeauthoryear{Ronnow-Rasmussen}{Ronnow-Rasmussen}{2015}]{handbook-intrinsic}
Ronnow-Rasmussen, T. (2015).
\newblock Intrisic and extrinsic value.
\newblock In I.~Hirose and J.~Olson (Eds.), {\em {The Oxford Handbook of Value
  Theory}}, Chapter~2, pp.\  29--43. Oxford University Press.

\bibitem[\protect\citeauthoryear{Sambasivan, Kapania, Highfill, Akrong,
  Paritosh, and Aroyo}{Sambasivan et~al.}{2021}]{data-cascades}
Sambasivan, N., S.~Kapania, H.~Highfill, D.~Akrong, P.~K. Paritosh, and
  L.~Aroyo (2021).
\newblock “everyone wants to do the model work, not the data work”: Data
  cascades in high-stakes ai.
\newblock {\em Proceedings of the 2021 CHI Conference on Human Factors in
  Computing Systems\/}.

\bibitem[\protect\citeauthoryear{Scao, Wang, Hesslow, Saulnier, Bekman, Bari,
  Biderman, ElSahar, Phang, Press, Raffel, Sanh, Shen, Sutawika, Tae, Yong,
  Launay, and Beltagy}{Scao et~al.}{2022}]{bigscience-architecture}
Scao, T.~L., T.~Wang, D.~Hesslow, L.~Saulnier, S.~Bekman, S.~Bari, S.~R.
  Biderman, H.~ElSahar, J.~Phang, O.~Press, C.~Raffel, V.~Sanh, S.~Shen, L.~A.
  Sutawika, J.~Tae, Z.~X. Yong, J.~Launay, and I.~Beltagy (2022).
\newblock What language model to train if you have one million gpu hours?

\bibitem[\protect\citeauthoryear{Scheuerman, Hanna, and Denton}{Scheuerman
  et~al.}{2021}]{dataset-politics}
Scheuerman, M.~K., A.~Hanna, and E.~Denton (2021, oct).
\newblock Do datasets have politics? disciplinary values in computer vision
  dataset development.
\newblock ~{\em 5\/}(CSCW2).

\bibitem[\protect\citeauthoryear{Srivastava, Rastogi, Rao, Shoeb, Abid, Fisch,
  Brown, Santoro, Gupta, Garriga-Alonso, Kluska, Lewkowycz, Agarwal, Power,
  Ray, Warstadt, Kocurek, Safaya, Tazarv, Xiang, Parrish, Nie, Hussain, Askell,
  Dsouza, Rahane, Iyer, Andreassen, Santilli, Stuhlmuller, Dai, La, Lampinen,
  Zou, Jiang, Chen, Vuong, Gupta, Gottardi, Norelli, Venkatesh, Gholamidavoodi,
  Tabassum, Menezes, Kirubarajan, Mullokandov, Sabharwal, Herrick, Efrat,
  Erdem, Karakacs, Roberts, Loe, Zoph, Bojanowski, Ozyurt, Hedayatnia,
  Neyshabur, Inden, Stein, Ekmekci, Lin, Howald, Diao, Dour, Stinson, Argueta,
  Ram'irez, Singh, Rathkopf, Meng, Baral, Wu, Callison-Burch, Waites, Voigt,
  Manning, Potts, Ramirez, Rivera, Siro, Raffel, Ashcraft, Garbacea, Sileo,
  Garrette, Hendrycks, Kilman, Roth, Freeman, Khashabi, Levy, Gonz'alez,
  Hernandez, Chen, Ippolito, Gilboa, Dohan, Drakard, Jurgens, Datta, Ganguli,
  Emelin, Kleyko, Yuret, Chen, Tam, Hupkes, Misra, Buzan, Mollo, Yang, Lee,
  Shutova, Cubuk, Segal, Hagerman, Barnes, Donoway, Pavlick, Rodol{\`a}, Lam,
  Chu, Tang, Erdem, Chang, Chi, Dyer, Jerzak, Kim, Manyasi, Zheltonozhskii,
  Xia, Siar, Mart'inez-Plumed, Happ'e, Chollet, Rong, Mishra, Winata, de~Melo,
  Kruszewski, Parascandolo, Mariani, Wang, Jaimovitch-L'opez, Betz, Gur-Ari,
  Galijasevic, Kim, Rashkin, Hajishirzi, Mehta, Bogar, Shevlin, Sch{\"u}tze,
  Yakura, Zhang, Wong, Ng, Noble, Jumelet, Geissinger, Kernion, Hilton, Lee,
  Fisac, Simon, Koppel, Zheng, Zou, Koco'n, Thompson, Kaplan, Radom,
  Sohl-Dickstein, Phang, Wei, Yosinski, Novikova, Bosscher, Marsh, Kim, Taal,
  Engel, Alabi, Xu, Song, Tang, Waweru, Burden, Miller, Balis, Berant,
  Frohberg, Rozen, Hern{\'a}ndez-Orallo, Boudeman, Jones, Tenenbaum, Rule,
  Chua, Kanclerz, Livescu, Krauth, Gopalakrishnan, Ignatyeva, Markert, Dhole,
  Gimpel, Omondi, Mathewson, Chiafullo, Shkaruta, Shridhar, McDonell,
  Richardson, Reynolds, Gao, Zhang, Dugan, Qin, Contreras-Ochando, Morency,
  Moschella, Lam, Noble, Schmidt, He, Col'on, Metz, cSenel, Bosma, Sap, ter
  Hoeve, Andrea, Farooqi, Faruqui, Mazeika, Baturan, Marelli, Maru, Quintana,
  Tolkiehn, Giulianelli, Lewis, Potthast, Leavitt, Hagen, Schubert,
  Baitemirova, Arnaud, McElrath, Yee, Cohen, Gu, Ivanitskiy, Starritt, Strube,
  Swkedrowski, Bevilacqua, Yasunaga, Kale, Cain, Xu, Suzgun, Tiwari, Bansal,
  Aminnaseri, Geva, Gheini, MukundVarma, Peng, Chi, Lee, Krakover, Cameron,
  Roberts, Doiron, Nangia, Deckers, Muennighoff, Keskar, Iyer, Constant,
  Fiedel, Wen, Zhang, Agha, Elbaghdadi, Levy, Evans, Casares, Doshi, Fung,
  Liang, Vicol, Alipoormolabashi, Liao, Liang, Chang, Eckersley, Htut, Hwang,
  Milkowski, Patil, Pezeshkpour, Oli, Mei, LYU, Chen, Banjade, Rudolph,
  Gabriel, Habacker, Delgado, Milli{\`e}re, Garg, Barnes, Saurous, Arakawa,
  Raymaekers, Frank, Sikand, Novak, Sitelew, Bras, Liu, Jacobs, Zhang,
  Salakhutdinov, Chi, Lee, Stovall, Teehan, Yang, Singh, Mohammad, Anand,
  Dillavou, Shleifer, Wiseman, Gruetter, Bowman, Schoenholz, Han, Kwatra, Rous,
  Ghazarian, Ghosh, Casey, Bischoff, Gehrmann, Schuster, Sadeghi, Hamdan, Zhou,
  Srivastava, Shi, Singh, Asaadi, Gu, Pachchigar, Toshniwal, Upadhyay, Debnath,
  Shakeri, Thormeyer, Melzi, Reddy, Makini, hwan Lee, Torene, Hatwar, Dehaene,
  Divic, Ermon, Biderman, Lin, Prasad, Piantadosi, Shieber, Misherghi,
  Kiritchenko, Mishra, Linzen, Schuster, Li, Yu, Ali, Hashimoto, Wu, Desbordes,
  Rothschild, Phan, Wang, Nkinyili, Schick, Kornev, Telleen-Lawton, Tunduny,
  Gerstenberg, Chang, Neeraj, Khot, Shultz, Shaham, Misra, Demberg, Nyamai,
  Raunak, Ramasesh, Prabhu, Padmakumar, Srikumar, Fedus, Saunders, Zhang,
  Vossen, Ren, Tong, Wu, Shen, Yaghoobzadeh, Lakretz, Song, Bahri, Choi, Yang,
  Hao, Chen, Belinkov, Hou, Hou, Bai, Seid, Xinran, Zhao, Wang, Wang, Wang, Wu,
  Singh, and Shaham}{Srivastava et~al.}{2022}]{bigbench}
Srivastava, A., A.~Rastogi, A.~B. Rao, A.~A.~M. Shoeb, A.~Abid, A.~Fisch, A.~R.
  Brown, A.~Santoro, A.~Gupta, A.~Garriga-Alonso, A.~Kluska, A.~Lewkowycz,
  A.~Agarwal, A.~Power, A.~Ray, A.~Warstadt, A.~W. Kocurek, A.~Safaya,
  A.~Tazarv, A.~Xiang, A.~Parrish, A.~Nie, A.~Hussain, A.~Askell, A.~Dsouza,
  A.~A. Rahane, A.~S. Iyer, A.~J. Andreassen, A.~Santilli, A.~Stuhlmuller,
  A.~M. Dai, A.~D. La, A.~K. Lampinen, A.~Zou, A.~Jiang, A.~Chen, A.~Vuong,
  A.~Gupta, A.~Gottardi, A.~Norelli, A.~Venkatesh, A.~Gholamidavoodi,
  A.~Tabassum, A.~Menezes, A.~Kirubarajan, A.~Mullokandov, A.~Sabharwal,
  A.~Herrick, A.~Efrat, A.~Erdem, A.~Karakacs, B.~R. Roberts, B.~S. Loe,
  B.~Zoph, B.~Bojanowski, B.~Ozyurt, B.~Hedayatnia, B.~Neyshabur, B.~Inden,
  B.~Stein, B.~Ekmekci, B.~Y. Lin, B.~S. Howald, C.~Diao, C.~Dour, C.~Stinson,
  C.~Argueta, C.~F. Ram'irez, C.~Singh, C.~Rathkopf, C.~Meng, C.~Baral, C.~Wu,
  C.~Callison-Burch, C.~Waites, C.~Voigt, C.~D. Manning, C.~Potts, C.~T.
  Ramirez, C.~Rivera, C.~Siro, C.~Raffel, C.~Ashcraft, C.~Garbacea, D.~Sileo,
  D.~H. Garrette, D.~Hendrycks, D.~Kilman, D.~Roth, D.~Freeman, D.~Khashabi,
  D.~Levy, D.~Gonz'alez, D.~Hernandez, D.~Chen, D.~Ippolito, D.~Gilboa,
  D.~Dohan, D.~Drakard, D.~Jurgens, D.~Datta, D.~Ganguli, D.~Emelin, D.~Kleyko,
  D.~Yuret, D.~Chen, D.~Tam, D.~Hupkes, D.~Misra, D.~Buzan, D.~C. Mollo,
  D.~Yang, D.-H. Lee, E.~Shutova, E.~D. Cubuk, E.~Segal, E.~Hagerman,
  E.~Barnes, E.~P. Donoway, E.~Pavlick, E.~Rodol{\`a}, E.~F. Lam, E.~Chu,
  E.~Tang, E.~Erdem, E.~Chang, E.~A. Chi, E.~Dyer, E.~Jerzak, E.~Kim, E.~E.
  Manyasi, E.~Zheltonozhskii, F.~Xia, F.~Siar, F.~Mart'inez-Plumed, F.~Happ'e,
  F.~Chollet, F.~Rong, G.~Mishra, G.~I. Winata, G.~de~Melo, G.~Kruszewski,
  G.~Parascandolo, G.~Mariani, G.~Wang, G.~Jaimovitch-L'opez, G.~Betz,
  G.~Gur-Ari, H.~Galijasevic, H.~S. Kim, H.~Rashkin, H.~Hajishirzi, H.~Mehta,
  H.~Bogar, H.~Shevlin, H.~Sch{\"u}tze, H.~Yakura, H.~Zhang, H.~Wong, I.~A.-S.
  Ng, I.~Noble, J.~Jumelet, J.~Geissinger, J.~Kernion, J.~Hilton, J.~Lee, J.~F.
  Fisac, J.~B. Simon, J.~Koppel, J.~Zheng, J.~Zou, J.~Koco'n, J.~Thompson,
  J.~Kaplan, J.~Radom, J.~N. Sohl-Dickstein, J.~Phang, J.~Wei, J.~Yosinski,
  J.~Novikova, J.~Bosscher, J.~Marsh, J.~Kim, J.~Taal, J.~Engel, J.~O. Alabi,
  J.~Xu, J.~Song, J.~Tang, J.~W. Waweru, J.~Burden, J.~Miller, J.~U. Balis,
  J.~Berant, J.~Frohberg, J.~Rozen, J.~Hern{\'a}ndez-Orallo, J.~Boudeman,
  J.~Jones, J.~B. Tenenbaum, J.~S. Rule, J.~Chua, K.~Kanclerz, K.~Livescu,
  K.~Krauth, K.~Gopalakrishnan, K.~Ignatyeva, K.~Markert, K.~D. Dhole,
  K.~Gimpel, K.~O. Omondi, K.~W. Mathewson, K.~Chiafullo, K.~Shkaruta,
  K.~Shridhar, K.~McDonell, K.~Richardson, L.~Reynolds, L.~Gao, L.~Zhang,
  L.~Dugan, L.~Qin, L.~Contreras-Ochando, L.-P. Morency, L.~Moschella, L.~Lam,
  L.~Noble, L.~Schmidt, L.~He, L.~O. Col'on, L.~Metz, L.~K. cSenel, M.~Bosma,
  M.~Sap, M.~ter Hoeve, M.~Andrea, M.~S. Farooqi, M.~Faruqui, M.~Mazeika,
  M.~Baturan, M.~Marelli, M.~Maru, M.~Quintana, M.~Tolkiehn, M.~Giulianelli,
  M.~Lewis, M.~Potthast, M.~Leavitt, M.~Hagen, M.~Schubert, M.~Baitemirova,
  M.~Arnaud, M.~A. McElrath, M.~A. Yee, M.~Cohen, M.~Gu, M.~I. Ivanitskiy,
  M.~Starritt, M.~Strube, M.~Swkedrowski, M.~Bevilacqua, M.~Yasunaga, M.~Kale,
  M.~Cain, M.~Xu, M.~Suzgun, M.~Tiwari, M.~Bansal, M.~Aminnaseri, M.~Geva,
  M.~Gheini, T.~MukundVarma, N.~Peng, N.~Chi, N.~Lee, N.~G.-A. Krakover,
  N.~Cameron, N.~S. Roberts, N.~Doiron, N.~Nangia, N.~Deckers, N.~Muennighoff,
  N.~S. Keskar, N.~Iyer, N.~Constant, N.~Fiedel, N.~Wen, O.~Zhang, O.~Agha,
  O.~Elbaghdadi, O.~Levy, O.~Evans, P.~A.~M. Casares, P.~Doshi, P.~Fung, P.~P.
  Liang, P.~Vicol, P.~Alipoormolabashi, P.~Liao, P.~Liang, P.~W. Chang,
  P.~Eckersley, P.~M. Htut, P.-B. Hwang, P.~Milkowski, P.~S. Patil,
  P.~Pezeshkpour, P.~Oli, Q.~Mei, Q.~LYU, Q.~Chen, R.~Banjade, R.~E. Rudolph,
  R.~Gabriel, R.~Habacker, R.~R. Delgado, R.~Milli{\`e}re, R.~Garg, R.~Barnes,
  R.~A. Saurous, R.~Arakawa, R.~Raymaekers, R.~Frank, R.~Sikand, R.~Novak,
  R.~Sitelew, R.~L. Bras, R.~Liu, R.~Jacobs, R.~Zhang, R.~Salakhutdinov,
  R.~Chi, R.~Lee, R.~Stovall, R.~Teehan, R.~Yang, S.~J. Singh, S.~M. Mohammad,
  S.~Anand, S.~Dillavou, S.~Shleifer, S.~Wiseman, S.~Gruetter, S.~Bowman, S.~S.
  Schoenholz, S.~Han, S.~Kwatra, S.~A. Rous, S.~Ghazarian, S.~Ghosh, S.~Casey,
  S.~Bischoff, S.~Gehrmann, S.~Schuster, S.~Sadeghi, S.~S. Hamdan, S.~Zhou,
  S.~Srivastava, S.~Shi, S.~Singh, S.~Asaadi, S.~S. Gu, S.~Pachchigar,
  S.~Toshniwal, S.~Upadhyay, S.~Debnath, S.~Shakeri, S.~Thormeyer, S.~Melzi,
  S.~Reddy, S.~P. Makini, S.~hwan Lee, S.~B. Torene, S.~Hatwar, S.~Dehaene,
  S.~Divic, S.~Ermon, S.~R. Biderman, S.~C. Lin, S.~Prasad, S.~T. Piantadosi,
  S.~M. Shieber, S.~Misherghi, S.~Kiritchenko, S.~Mishra, T.~Linzen,
  T.~Schuster, T.~Li, T.~Yu, T.~A. Ali, T.~Hashimoto, T.-L. Wu, T.~Desbordes,
  T.~Rothschild, T.~Phan, T.~Wang, T.~Nkinyili, T.~Schick, T.~N. Kornev,
  T.~Telleen-Lawton, T.~Tunduny, T.~Gerstenberg, T.~Chang, T.~Neeraj, T.~Khot,
  T.~O. Shultz, U.~Shaham, V.~Misra, V.~Demberg, V.~Nyamai, V.~Raunak, V.~V.
  Ramasesh, V.~U. Prabhu, V.~Padmakumar, V.~Srikumar, W.~Fedus, W.~Saunders,
  W.~Zhang, W.~Vossen, X.~Ren, X.~F. Tong, X.~Wu, X.~Shen, Y.~Yaghoobzadeh,
  Y.~Lakretz, Y.~Song, Y.~Bahri, Y.~J. Choi, Y.~Yang, Y.~Hao, Y.~Chen,
  Y.~Belinkov, Y.~Hou, Y.~Hou, Y.~Bai, Z.~Seid, Z.~Xinran, Z.~Zhao, Z.~F. Wang,
  Z.~J. Wang, Z.~Wang, Z.~Wu, S.~Singh, and U.~Shaham (2022).
\newblock Beyond the imitation game: Quantifying and extrapolating the
  capabilities of language models.
\newblock {\em ArXiv\/}~{\em abs/2206.04615}.

\bibitem[\protect\citeauthoryear{Sullins}{Sullins}{2021}]{sullins-divide}
Sullins, J. (2021).
\newblock {Information Technology and Moral Values}.
\newblock In E.~N. Zalta (Ed.), {\em The {Stanford} Encyclopedia of
  Philosophy\/} ({S}pring 2021 ed.). Metaphysics Research Lab, Stanford
  University.

\bibitem[\protect\citeauthoryear{Symons and Horner}{Symons and
  Horner}{2014}]{Symons2014-software-intensive}
Symons, J. and J.~Horner (2014).
\newblock Software intensive science.
\newblock {\em Philosophy and Technology\/}~{\em 27\/}(3), 461--477.

\bibitem[\protect\citeauthoryear{Wang and Barabási}{Wang and
  Barabási}{2021}]{wang_barabási_2021}
Wang, D. and A.-L. Barabási (2021).
\newblock {\em The Science of Science}.
\newblock Cambridge University Press.

\bibitem[\protect\citeauthoryear{Winner}{Winner}{1980}]{artifact-politics}
Winner, L. (1980).
\newblock Do artifacts have politics?
\newblock {\em Daedalus\/}~{\em 109\/}(1), 121--136.

\bibitem[\protect\citeauthoryear{Young, Katell, and Krafft}{Young
  et~al.}{2022}]{facct-corporate-capture}
Young, M., M.~Katell, and P.~Krafft (2022).
\newblock Confronting power and corporate capture at the facct conference.
\newblock In {\em 2022 ACM Conference on Fairness, Accountability, and
  Transparency}, FAccT '22, New York, NY, USA, pp.\  1375–1386. Association
  for Computing Machinery.

\end{thebibliography}

\end{document}